# Unconventional Resistivity Scaling in Topological Semimetal CoSi

Shang-Wei Lien[1,10], Ion Garate*[2,10], Utkarsh Bajpai[3], Cheng-Yi Huang[4], Chuang-Han Hsu[5], Yi-Hsin Tu[1], Nicholas A. Lanzillo[3], Arun Bansil[4], Tay-Rong Chang[1,6,7], Gengchiau Liang*[8], Hsin Lin*[5], Ching-Tzu Chen*[9]

[1] Department of Physics, National Cheng Kung University, Tainan 701, Taiwan

[2] Département de Physique, Institut Quantique and Regroupement Québécois sur les Matériaux de Pointe, Université de Sherbrooke, Sherbrooke, Québec, Canada J1K 2R1

[3] IBM Research, 257 Fuller Road, Albany, NY 12203, USA

[4] Department of Physics, Northeastern University, Boston, Massachusetts 02115, USA

[5] Institute of Physics, Academia Sinica, Taipei 11529, Taiwan

[6] Center for Quantum Frontiers of Research & Technology (QFort), Tainan 701, Taiwan

[7] Physics Division, National Center for Theoretical Sciences, Hsinchu, Taiwan

[8] Department of Electrical and Computer Engineering, College of Design and Engineering, National University of Singapore, Singapore, Singapore

[9] IBM T.J. Watson Research Center, Yorktown Heights, NY 10598, USA

[10] These authors contributed equally.

Nontrivial band topologies in semimetals lead to robust surface states that can contribute dominantly to the total conduction. This may result in reduced resistivity with decreasing feature size contrary to conventional metals, which may highly impact the semiconductor industry. Here we study the resistivity scaling of a representative topological semimetal CoSi using realistic band structures and Green's function methods. We show that there exists a critical thickness $d_c$ dividing different scaling trends. Above $d_c$, when the defect density is low such that surface conduction dominates, resistivity reduces with decreasing thickness; when the defect density is high such that bulk conduction dominates, resistivity increases as in conventional metals. Below $d_c$, the persistent remnants of the surface states give rise to decreasing resistivity down to the ultrathin limit, unlike in topological insulators. The observed CoSi scaling can apply to broad classes of topological semimetals, providing guidelines for materials screening and engineering. Our study shows that topological semimetals bear the potential of overcoming the resistivity scaling challenges in back-end-of-line interconnect applications.

* Email: ion.garate@usherbrooke.ca; elelg@nus.edu.sg; nilnish@gmail.com; cchen3@us.ibm.com

# I. INTRODUCTION

Semiconductor industry currently faces a roadblock: as integrated circuits become smaller, the ever-increasing electrical resistivity of standard Cu interconnects hampers the circuit power-performance despite the downscaling of transistors. The search for alternative interconnect materials is actively underway[1,2]. Among them, topological semimetals have recently been identified as potentially promising[1,3–5].

Topological semimetals comprise unconventional materials in which the conduction and valence bands touch at discrete pairs of nodal points in the Brillouin zone near the Fermi level[6–9]. The nontrivial bulk-band topology near the band-crossings yields robust surface states that connect between the nodes, forming open Fermi arcs. Both surface and bulk states contribute to rich transport properties[10–12]. In the simplest Weyl semimetal comprising two Weyl nodes, carrier transport via the Fermi-arc states resembles that of the edge-states in a stack of 2D quantum anomalous Hall insulators[13], yielding size-independent contributions to total conductivity at macroscopic scales[14]. In a nonmagnetic Weyl semimetal, if the relaxation lengths between time-reversed partners exceed the sample thickness, similar quantum Hall physics would hold, as if there were two independent Fermi-arc conduction channels[14].

Furthermore, the increasing surface-to-bulk ratio with reduced sample dimensions should lead to *decreased* resistivity due to Fermi-arc transport[15]. Indeed, measurements of NbAs and CoSi appear to support this conjecture: resistivity of ~100 nm thick NbAs nanobelts drops ≥ 10-fold below its bulk value to ~1 $\mu\Omega$-cm[16] (lower than bulk Cu resistivity), while that of CoSi nanowires decreases with reduced diameters to ≤ 1/5 of its bulk value[17]. This differs sharply from conventional metals (e.g. Cu) whose resistivity *increases* with reduced dimensions due to carrier scattering off surfaces, grain-boundaries, and so on[2]. The increase in wiring resistance has become a major challenge for advanced semiconductor technologies. In comparison, the resistivity reduction in nanoscale NbAs and CoSi may yield significant performance gains[1,3]. For example, if the low resistivity observed in the NbAs nanobelts[16] can persist down to the nanometer scale and can be observed in a wider class of materials compatible with silicon technologies, then replacing copper with topological semimetals for wiring up the transistors can improve the power-performance of integrated circuits by at least 10% to ~40%, equivalent to 1 to 4 technology-node generations of performance gain[3], promoting topological semimetals for beyond Cu interconnects among other applications[1,4,18,19]. It is thus imperative to understand the underlying physics and generality of such a resistivity scaling for a broad class of topological semimetals. To this end, we report for the first time a rigorous study using a combined first-principles and analytical modeling approach for a representative topological semimetal, the chiral multifermion semimetal CoSi, a material that can be readily integrated in the silicon technology. Our work elucidates the various resistivity scaling regimes in thin films with thickness ranging from 2 to ~50 unit cells and reveals the conditions under which novel scaling reverts to conventional metallic scaling.

## II. ELECTRONIC STRUCTURE OF CoSi

CoSi is nonmagnetic with a chiral cubic structure (space group # 198) that breaks the inversion symmetry (Fig. 1a). In the absence of spin-orbit coupling, density-functional theory (DFT) derived bulk band structure reveals a three-fold degenerate band-crossing at Γ and a four-fold degenerate band-crossing at R (Fig. 1c), with Chern numbers +2 and −2 respectively[20,21]. Consequently, two Fermi arcs (per spin) connect the $\bar{\Gamma}$ and $\bar{R}$ points, spanning the entire projected Brillouin Zone (BZ) along the (100) and equivalent surfaces (Fig. 1d), as confirmed by the angle-resolved photoemission spectroscopy[22–24]. In the presence of spin-orbit coupling (SOC), each of the doubly degenerate Fermi arcs is split into two. The total number of Fermi arcs extending from near $\bar{\Gamma}$ to near $\bar{R}$ remains the same (4). Therefore, the surface-state contributions to transport would not change qualitatively. Experimentally, since the SOC induced energy splitting is smaller than other broadenings in the CoSi samples, ARPES cannot resolve the Fermi-arc splitting[22–24]. Thus, within the experimental tolerance, the electronic states near the Fermi level are well described by a model without SOC.

## III. RESISTANCE-AREA SCALING IN CoSi SLABS WITH LINE DEFECTS

We first study the density-functional-theory (DFT) informed quantum transport of electrons along the [001] direction in (100)-oriented CoSi slabs of varying thickness, using a 2-terminal device configuration (Fig 2a) with and without line defects. The slab structure is finite along $y$ and infinite along $z$, and the transport direction points along $x$. The line defect in the form of a surface notch extends from $z = -\infty$ to $z = +\infty$ while preserving translational symmetry along $z$. The self-consistent single-particle Kohn-Sham (KS) Hamiltonian $H_{KS}$ of the scattering region in Fig. 2a is evaluated using the QuantumATK package[25]. The total conductance per unit area is calculated using the Nonequilibrium Green's Function (NEGF) technique[26] (see App. A).

Figure 2b depicts the Fermi surface of a 40 atomic-layer (AL) CoSi slab (thickness $d = 36.05$ Å), showing that conduction at the Fermi level $E_F$ predominantly comes from the Fermi arcs (denoted by red and blue lines) permeating throughout the projected BZ, while the bulk states (in gray) concentrate near [0,0] and [π, π]. Correspondingly, the local density of states (LDOS) near the surfaces is significantly higher than the LDOS in the bulk (see Supplementary Fig. 1a in [27]), and the thinner the slab, the larger the surface- to total-current ratio (see Supplementary Fig. 1b in [27]). Furthermore, the band structure along a linecut (e.g., L1 along $k_z = 0.52\pi/a$) in the 2D BZ shows that the left- and right-moving Fermi-arc states (S1 and S2) reside on *opposite* surfaces (see top panel of Fig. 2c), resembling the chiral edge states in quantum Hall insulators. Consequently, in CoSi slabs with line defects that preserve the translational invariance along $z$, if the slab is thick enough such that the top and bottom surface states do not overlap, backscattering between them is *negligible*. This is confirmed in the $k$-resolved transmission $T(k_z)$ (Fig. 2d), where the transmission remains intact between $k_z \sim 0.4\pi/a$ and $\sim 0.7\pi/a$ despite the surface defect.

Between $k_z \sim 0.2\pi/a$ and $\sim 0.4\pi/a$, multiple Fermi-arc states participate in transport while the bulk conduction is negligible. At a fixed $k_z$, only an odd number of surface states exist per surface per spin. For example, along $k_z = 0.3\pi/a$ (L2 in Fig. 2b), there are three states on the top (S3, S5, and S7) and bottom (S4, S6, and S8) surfaces, respectively. Just as S1 and S2, the states related by a C2 rotation along the $z$-axis (e.g., S4 and S7) are located on opposite surfaces and cannot backscatter among each other. Nevertheless, S4 can backscatter to S8. Therefore, scattering off a line defect does reduce $T$ (e.g., from 3 to ~2 at $k_z = 0.3\pi/a$), but $T(k_z)$ remains larger than one because band topology guarantees an extra forward-moving surface mode for transport. This protected Fermi-arc transmission can be seen in CoSi slabs with various types of line defects (see Supplementary Fig. 2 in [27]).

Figure 2e shows how the Fermi-arc states impact the scaling of resistance-area (RA) product. The number of Fermi-arc channels remains nearly constant with decreasing film thickness and, at the nanoscale, surface conduction dominates over bulk conduction. Thus, when thickness is reduced, the conductance per unit cross-sectional area (G/A) *increases*, i.e., the slab RA *decreases* to well below the bulk RA (see the orange curve with filled circles). This sharply contrasts the scaling in conventional metals, e.g., Cu, where the bulk states carry the conduction and therefore $(RA)_{slab}/(RA)_{bulk} \sim 1$ in pristine films even at nanoscale (orange curve with open circles). The contrast is even more drastic when there is disorder: scattering of bulk carriers off surface defects in Cu yields an increasing RA with reduced thickness (black curve with open circles), while the protected transmission of Fermi-arc carriers in CoSi maintains the *decreasing* RA with reduced thickness (black curve with filled circles) despite some loss in transmission. We note that, in the presence of SOC, the split Fermi arcs still extend from near $\bar{\Gamma}$ to near $\bar{R}$, spanning a large phase space in the BZ. Thus, surface-state transport still dominates. Furthermore, at a fixed $k_z$, electrons of the split surface bands travel along the *same* direction, and the chiral edge-state like characteristic remains. Thus, the trend of decreasing RA with scaling holds regardless of SOC.

Translational invariance of the line defects ensures that electrons only scatter between the Fermi-arc surface states with the same $k_z$ (see Fig. 2b), resulting in a protected chiral Fermi-arc transport. However, in the presence of point defects, this is no longer true, and an enlarged phase space for scattering leads to a more complex scaling behavior.

## IV. RESISTIVITY SCALING IN CoSi SLABS WITH POINT DEFECTS

When there are point defects, the DFT-informed NEGF calculations become prohibitively expensive. Thus, we perform calculations within the tight-binding formalism by constructing Wannier functions that can reproduce the band structures generated by QuantumATK. This method enables us to extend the relaxation length and conductivity calculations for CoSi films from a few unit-cell to over 100-unit-cell thick. Furthermore, instead of using wave functions explicitly in the Fermi golden rule and Kubo's formula[28,29], we adopt Green's functions in the computation in the matrix form, which accounts for all matrix-element effects, including suppression of transitions between opposite (pseudo-)spins. This approach speeds up the convergence. It also allows us to introduce an energy broadening factor $\eta$ (~1 meV) as the simplest model for other scattering sources that reduce the lifetime of the entire system uniformly. Next, we introduce Co vacancies on the film surface (Fig. 3a), construct the corresponding impurity potential and T-matrix (see Apps. B and C), and calculate the conductivity of the slab using Kubo's formula (see Apps. F and G). Cobalt vacancies are used to model the surface defects because their formation energy is lower than Si vacancies. We focus on the impact of surface defects instead of bulk defects because Fermi-arc states dominate the transport of pristine CoSi thin films (see Supplementary Fig. 1 in [27]).

Figures 3b and 3c summarize the key results of this work: resistivity scaling of CoSi with film thickness ($\rho/\rho_0$ vs. $d$) and surface defect density $N$ (defined as the number of defects per unit cell of the slab), where $\rho_0$ denotes the resistivity of an infinite bulk. We observe that, when $N$ is small, $\rho/\rho_0$ decreases with decreasing thickness $d$, consistent with the RA scaling in slabs with a shallow notch (Fig. 2e). When $N$ is large, resistivity scales differently above (region $I$) and below (region $II$) a "critical" thickness $d_c$ of 6 unit cells (denoted by the dashed line). Next, we explain this scaling behavior with an analytical model.

### A. Resistivity scaling above the critical thickness

When $d > d_c$, well-differentiated surface and bulk channels conduct in parallel and hence $\rho/\rho_0 = \sigma_0/(\sigma_b + \sigma_s)$, where $\sigma_0 = 1/\rho_0$ is the conductivity of an infinite bulk and $\sigma_s$ ($\sigma_b$) is the average contribution from surface (bulk) states to the 3D total conductivity of the film[14]. In this regime, the $d$-dependence of $\rho$ follows from the competition between two opposite trends.

On the one hand, $\sigma_s$ grows as $d$ decreases because the number of Fermi-arcs is independent of $d$. Hence, for thinner samples, Fermi arcs make a higher average contribution to the 3D conductivity. Specifically, an analytical theory developed in the Supplemental Material [27] shows that

$$\sigma_s = \frac{2e^2}{\pi h} k_0 \frac{l_s}{d}, \qquad (1)$$

where $k_0 = \sqrt{2}\,\pi/a$ is the effective length of the Fermi arcs and $l_s$ is an effective scattering length for surface electrons. Above $d_c$, $l_s$ hardly varies with $d$ (see Supplementary Fig. 3 in [27]), which can be understood by a dimensional analysis argument[27]. Therefore, $\sigma_s$ scales roughly with $d^{-1}$.

On the other hand, $\sigma_b$ decreases as $d$ decreases (Fig. 4a). Since the bulk mean free path $l_0$ and the bulk density of states in region $I$ are approximately independent of $d$ (see [27] Sec. S1 and Supplementary Fig. 4), the observed behavior can be attributed to the Fuchs-Sondheimer (FS) mechanism[30], namely, the diffusive scattering of bulk electrons from the surfaces. Indeed, the calculated $\sigma_b$ vs $d$ curves fit well to the FS theory (see Fig. 4b and App. I).

Which of these two competing trends dominates in the overall resistivity scaling depends on the $\sigma_s/\sigma_b$ ratio. Combining Eq. (1) with the FS theory for $\sigma_b$, we have

$$\frac{\sigma_s}{\sigma_b} \sim \frac{(e^2/h)k_0}{\sigma_0} \frac{l_s l_0}{d^2 \ln(l_0/d)} \sim \frac{k_0 l_s}{(k_F d)^2} \frac{1}{\ln(l_0/d)} \qquad (2)$$

for $l_0 \gg d$, where $k_F \ll 1/a$ is the maximal Fermi wave vector of the bulk crystal for the dominant Fermi pocket. Thus, the main factors enhancing the relative contribution of the surface states to the average 3D conductivity include: (i) long Fermi arcs ($k_0 \gg k_F$), (ii) long surface scattering lengths ($l_s \gg d$), and (iii) small film thickness ($d \lesssim k_F^{-1}$). When $N = 0$, we find $\sigma_s/\sigma_b \gtrsim 1$ for the range of $d$ studied (see Supplementary Fig. 5 in [27]), mainly due to $k_0 \gg k_F$ and $l_s \gg d$. In this case, the surface contribution is dominant. Therefore $\rho/\rho_0$ decreases as $d$ decreases. As $N$ increases, the reduction of $l_s$ and $l_0$ results in a decrease of $\sigma_s/\sigma_b$ (see Eq. (2) above and Supplementary Fig. 5 in [27]). For sufficiently large $N$, $\sigma_s/\sigma_b$ drops appreciably below unity, and the bulk contribution becomes dominant, resulting in the sign change of the slope of $\rho/\rho_0$ vs $d$ (Figs. 3b and 3c).

## B. Resistivity scaling in the ultrathin limit

When $d < d_c$, according to Fig. 5, the only states present at the Fermi level are the remnants of Fermi arcs whose wave functions now spread throughout the entire film volume. For these states, the 3D conductivity roughly scales as $\sigma \propto \tau/d$, where $\tau$ is a scattering time and the $1/d$ factor originates from the fact that the number of surface bands does not depend on the film thickness. From Matthiessen's rule, the scattering rate is a sum of the rates due to surface vacancies and other sources, i.e., $\tau^{-1} \sim a\,n_{\text{imp}} + b\,\eta$, where $n_{\text{imp}} = N/d$ is the volume impurity density and $a, b$ are approximately constants. Accordingly, $\rho/\rho_0 = \sigma_0/\sigma \propto a\,N + b\,d\,\eta$ is roughly a linear function of $d$, with an intercept proportional to $N$ and a slope independent of $N$. This simple functional form fits qualitatively well to the numerical results in region *II* of Fig. 3b (see Supplementary Fig. 6 in [27]). In summary, the Fermi-arc contribution to the 3D conductivity remains remarkably robust down to the thinnest films and thereby enables a favorable resistivity scaling.

## C. Resistivity at the critical thickness

There are two noticeable features in Fig. 3b associated with $d = d_c$. First, for large $N$, the slope of $\rho/\rho_0$ vs. $d$ reverses when crossing over from above to below $d_c$, which is explained in the preceding discussion. Second, there is a resistivity kink at $d_c$ because at $d = d_c$, the bulk conduction band around $R$ starts to emerge at the Fermi level, giving rise to a van Hove singularity (Figs. 5a-c).

On one hand, the van Hove singularity contributes to more conduction channels, which would increase the total conductivity. On the other hand, the appearance of bulk states at the Fermi level reduces the lifetime for the remnants of the Fermi-arc states, which would decrease the total conductivity. Since the contribution from the emerging bulk states to conductivity is still small at $d \simeq d_c$ (the group velocity of the bulk states at the Fermi level being particularly small at $d = d_c$), the increased phase space for scattering is the dominant factor and gives rise to an upward kink in resistivity at $d_c$.

## V. DISCUSSION AND CONCLUSION

In summary, we have introduced a method that extends the first-principles based transport calculations to thin films of ~100 unit-cell thickness (an over tenfold improvement over the state-of-the-art in system size), validated against our analytical framework. Using this method, we have obtained the resistivity scaling for a silicon CMOS-compatible topological semimetal CoSi from 2 unit-cell (< 1nm) to ~50 unit-cell thickness, exposing a full range of scaling behavior and placing the prospect of using topological semimetals for ultra-scaled interconnects on a firm theoretical basis.

We conclude with a discussion on the resistivity scaling of general topological semimetals, focusing on materials with well-separated bulk and Fermi-arc surfaces states. First, we note the similarities and the differences between the $\rho$ vs. $d$ scaling of the chiral CoSi in Fig. 3b and that of the Weyl semimetal toy model in Fig. 4 of Ref. [14]. In symmetry-protected multifermion semimetals[31], such as CoSi, each Weyl node is its own time-reversed partner. Thus, the pair of time-reversed subsystems are strongly coupled. Consequently, instead of a two-step drop in resistivity with a plateau in between that manifests the two channels of anomalous Hall current of the two decoupled time-reversed subsystems in nonmagnetic Weyl semimetals[14], we see a one-step resistivity drop caused by the dominance of surface conduction over bulk conduction with scaling.

In short, in clean samples with *low* defect densities $N$ and thicknesses above $d_c$, the monotonically decreasing resistivity in CoSi is a general feature for nonmagnetic topological semimetals with strongly coupled time-reversed partners, such as chiral multifermion semimetals[20–24] and Dirac semimetals with Fermi arcs[31,32]. It is also a general feature for magnetic Weyl semimetals with broken time-reversal symmetry[34–36]. In contrast, the two-step resistivity decrease shown in Ref. [14] is a general feature for nonmagnetic topological semimetals with well separated time-reversed subsystems, such as TaAs[37,38].

On the other hand, in samples with *high N* and thicknesses above $d_c$, the increased resistivity with scaling due to the FS mechanism is a universal feature for most topological semimetals. This is broadly applicable to materials in which the bulk conduction dominates over surface conduction ($\sigma_b/\sigma_s > 1$), as in conventional metals.

Below $d_c$, where the surface-bulk separation is no longer valid, the decreased resistivity with scaling in CoSi regardless of $N$ is a general feature for topological semimetals with conducting surface states down to the ultrathin limit (see Supplementary Fig. 7 in [27]), in sharp contrast to prototypical topological insulators, such as $Bi_2Se_3$, where the topological surface states are gapped out. This property is highly desirable for nanometer-scale interconnects and can be found in topological materials with a sufficiently small overlap between the top- and bottom-surface Fermi arcs in the BZ. Examples include topological semimetals with chiral crystalline structures (e.g., CoSi, RhSi, etc.)[20,21,39] and Weyl semimetals with asymmetric top and bottom surface terminations, comprising different types of atoms (e.g., TaAs, NbAs)[37,38,40–42].

Besides the small coupling between the asymmetric top- and bottom-surface Fermi arcs, other desirable properties that can promote surface conduction include long Fermi arcs (e.g., CoSi, RhSi, AlPt, etc.)[23,24,39], many well-separated Weyl nodes, well-separated time-reversed surface-state partners to enhance the surface scattering lengths (e.g., TaAs), and conduction via chiral-edge-state like Fermi arcs where carriers traveling in opposite directions locate on opposite surfaces to suppress backscattering (e.g., magnetic Weyl semimetals). These serve as the guiding principles for screening topological interconnect materials. Whether or not the abovementioned factors can be satisfied simultaneously is a direction for further research. Detailed analyses of the electron-phonon scattering[43–45] on resistivity scaling in nanoscale topological semimetals would also be highly valuable.

## ACKNOWLEDGEMENTS

I.G. acknowledges financial support from the Natural Sciences and Engineering Research Council of Canada (Grant No. RGPIN- 2018- 05385), and the Fonds de Recherche du Québec Nature et

Technologies. G.L. acknowledges the support under grant number MOE-2019-T2-2-215 and FRC-A-8000194-01-00. H.L. acknowledges the support from the Ministry of Science and Technology (MOST) in Taiwan under grant number MOST 109-2112-M-001-014-MY3. The work at Northeastern University was supported by the Air Force Office of Scientific Research under award number FA9550-20-1-0322 and it benefited from the computational resources of Northeastern University's Advanced Scientific Computation Center (ASCC) and the Discovery Cluster. C.-T.C. acknowledges the fruitful discussions with R. Sundararaman, S. Kumar, J. Cha, C. Hinkle, and P. O. Sukhachov.

## APPENDIX A: DFT-INFORMED QUANTUM TRANSPORT CALCULATION FOR CoSi SLABS WITH 1D LINE DEFECTS

The self-consistent single-particle Kohn-Sham (KS) Hamiltonian $H_{KS}$ of the scattering region in Fig. 2a is evaluated using Synopsys' QuantumATK package, where we employ double-zeta polarized localized orbitals as the basis set and the Perdew-Burke-Enzerhof (PBE) generalized gradient approximation (GGA) for the exchange correlation function [46]. The slab structures have a finite thickness along the *y*-direction and periodic boundary conditions in the *x-z*-plane with a 9 × 9 *k*-point grid that has been checked for convergence. In addition, all structures have been relaxed such that the force on every atom is less that 5 meV/ Å.

The zero-temperature total conductance per unit length ($G$) of the scattering region in Fig. 2a in the zero-bias limit ($eV_{bias} \ll E_F$) can be calculated using the Nonequilibrium Green's Function (NEGF) technique:

$$G(E_F) = \frac{G_0}{2\pi} \int dk_z \, T(k_z, E_F),  \qquad (3)$$

where $G_0 = 2e^2/h$ is the conductance quantum and $T(k_z, E) = Tr\,(G^r \Gamma_L G^a \Gamma_R)$ *is the* $k_z$-resolved transmission. Here, $G^r(k_z, E) = [E + i\eta - H_{KS}(k_z) - \Sigma(k_z, E)]^{-1} = [G^a(k_z, E)]^\dagger$ is the retarded Green's function, $\Sigma(k_z, E)$ is the self-energy matrix of the left (L) and right (R) leads shown in Fig. 2a, and $G^a(k_z, E)$ is the advanced Green's function. Lastly, $\Gamma_\alpha = i(\Sigma_\alpha - \Sigma_\alpha^\dagger)$ is the level-broadening matrix of lead-$\alpha$ ($\alpha = $ L, R).

## APPENDIX B: MODEL HAMILTONIANS OF CoSi BULK, SLABS AND VACANCIES

The Wannier-type Hamiltonian within the tight-binding (TB) formalism of the CoSi pristine bulk system is derived from the first-principles calculations using the Synopsys QuantumATK package with the *d*-orbitals of Co and *p*-orbitals of Si atoms, respectively. In Fig. 1d, we calculate the spectral weight of the surface and bulk states of a semi-infinite bulk using the iterative Green's function, based on the method of cyclic reduction of block-tridiagonal matrices introduced in Ref. [47].

The CoSi slab models with finite thickness $d$ are built with the TB Hamiltonians truncated in the real space along the thickness direction ($y$). The Co vacancies on the top and bottom surface (Fig. 3a) are modelled by adding an impurity potential $V$ that removes all hopping terms of the Hamiltonian to and from the vacancy sites and setting the on-site energy of the fictitious vacancy atoms to $U_0$. The value of $U_0$ determines the energy of the impurity level, which is unphysical for vacancies. We set $U_0 = 10$ eV above the Fermi level and find that such a high-energy spurious impurity level has negligible effect on the resistivity scaling, as resistivity reflects only the low-energy physics near the Fermi level.

Note that since we consider non-magnetic defects and ignore the spin-orbit coupling, intermixing between spin-up and spin-down electrons is forbidden. When spin-orbit coupling is included, scattering between opposite pseudo-spins would similarly be suppressed. Thus, we expect the resistivity scaling to follow qualitatively the same trend as shown in Figure 3b.

**APPENDIX C: T-MATRIX CALCULATION**

The impurity potential in momentum space $V_q$ can be obtained by the Fourier transform of the impurity potential in real space. By summing up all diagrams involving multiple scatterings off the impurity, the total scattering matrix $T(k, k')$ can be formulated as,

$$T(k, k') = V_{k-k'} + \frac{1}{N_{k''}} \sum_{k''} V_{k-k''} G^0_{k''}(E) T(k'', k') = V_{k-k'} \left[ 1 - \frac{1}{N_{k''}} \sum_{k''} G^0_{k''}(E) V_{k''-k'} \right]^{-1} \quad (4)$$

where $N_{k'(k'')}$ is the total number of $k'(k'')$ points; $G^0_{k'(k'')}(E)$ is the unfolded bare Green's function at $k'(k'')$ point for a given energy $E$. In this work, when T-matrix is involved, the sum over the momentum space goes up to the second Brillouin zone. Inclusion of the third zone has been tested in a few cases, and less than 10% difference is found.

**APPENDIX D: SCATTERING LENGTH CALCULATION BY THE FERMI GOLDEN RULE**

In the Supplemental Material Sec. S1, we derive the Fermi golden rule expressions for the scattering lengths (Eq. S29). For first-principles calculations, it is advantageous to use the Green's function instead of wave functions. As a result, Eq. (S29) can be rewritten as

$$\frac{1}{l_{s\bar{s}}} = \frac{2\pi}{a} \frac{1}{N_k} \sum_{k \in s} \text{Im} \left( \text{Tr} \left( G^0_k \cdot \frac{-1}{\pi} N \cdot \text{Im}(R_{\bar{s}}(k)) \right) \right) \quad (5.1)$$

$$\frac{1}{l_{sb}} = \frac{2\pi}{a} \frac{1}{N_k} \sum_{k \in s} \text{Im} \left( \text{Tr} \left( G^0_k \cdot \frac{-1}{\pi} N \cdot \text{Im}(R_b(k)) \right) \right), \quad (5.2)$$

where "Im" stands for the imaginary part,

$$R_{\bar{s}}(k) = \frac{1}{N_{k'}} \sum_{k' \in \bar{s}} T^\dagger(k,k') G^0_{k'} T(k',k) \tag{6.1}$$

$$R_b(k) = \frac{1}{N_{k'}} \sum_{k' \in b} T^\dagger(k,k') G^0_{k'} T(k',k), \tag{6.2}$$

$a$ is the lattice constant, $N_{k(k')}$ is the total number of $k(k')$ points in the summation, $N$ is the number of defects per unit cell of the slab, and the indices of $s, \bar{s}$ in Eq. (5.1) (or $s, b$ in Eq. (5.2)) denote the mean scattering length between time-reversed surface states (or between surface and bulk states). The separation of the surface and bulk regions in the Brillouin zone is described in the next section. Last, in Eq. (5.1) and Eq (5.2), the trace Tr is taken over the electronic orbitals in the Hamiltonian matrix.

## APPENDIX E: SEPARATION OF SURFACE AND BULK STATES

Because the bulk states can have nonzero spectral weights at the surface and the surface states can mix with the bulk states, in general, a rigorous separation between two cannot be done for finite slabs. Instead, we adopt a practical method to separate the surface and bulk states in the momentum space. The main concept is that any states that appear in the bulk energy gap region (which is forbidden under the 3D periodic condition) must involve surface terminations and have an exponentially decaying wave function amplitude in the bulk, and therefore, those states are defined as the surface states. At each in-plane $(k_z, k_x)$ point, we first identify the conduction-band minimum energy $E_c(k_z, k_x)$ and the valence-band maximum $E_v(k_z, k_x)$ of the *three-dimensional (3D)* CoSi bulk over all $k_y$, where y is the out-of-plane direction. We then compute the eigen-energies for each $(k_z, k_x)$ of a *two-dimensional (2D)* CoSi slab, $E_{2D}^n(k_z, k_x)$, where n denotes the band index. If $E_{2D}^n(k_z, k_x)$ exists in the bulk band gap, i.e., $E_v(k_z, k_x) < E_{2D}^n(k_z, k_x) < E_c(k_z, k_x)$, it is labeled as a surface state; otherwise, it is labeled as a bulk state. Since we only focus on the states near the Fermi level, our method divides the Brillouin zone into two regions: the surface states (inside the bulk band gap) and the bulk states, without any overlap. In some materials, there may exist surface resonances that intermix with the bulk bands. Nevertheless, in CoSi above the critical thickness, the effects of such surface resonances (if any) are negligible, as manifested by the constant density of the bulk states shown in Supplementary Fig. 4b in [27].

## APPENDIX F: CONDUCTIVITY CALCULATIONS WITH KUBO'S FORMULA

The electric conductivity of a 2D slab system can be computed by Kubo's formula[29]:

$$\sigma^{2D} \equiv \sigma_{xx}^{2D} = 2\frac{e^2}{h} \int_{\vec{k}} \frac{d^2k}{(2\pi)^2} \operatorname{Re} \operatorname{Tr}\left[\frac{\partial H}{\partial k_x}\left(G^a(\vec{k}, E_F) - G^r(\vec{k}, E_F)\right) \frac{\partial H}{\partial k_x} G^r(\vec{k}, E_F)\right], \tag{7}$$

where "Re" stands for the real part, the factor of two is the spin degeneracy, $H_{\vec{k}}$ is the TB Hamiltonian of the CoSi slab, $\frac{\partial H}{\partial k_x}$ is the x-direction velocity operator matrix, the trace is taken over the electronic orbitals of the Hamiltonian, and $G^r$ ($G^a$) is the retarded (advanced) Green's function,

$$G^r(\vec{k}, E) = \frac{1}{E + i\eta - H_{\vec{k}} - \Sigma(k)} \tag{8}$$

$$G^a(\vec{k}, E) = G^r(\vec{k}, E)^\dagger. \tag{9}$$

Here, $\eta \sim 1$ meV is a constant that simulates the effect of scattering sources other than surface vacancies (e.g. bulk impurities, phonons, thermal broadening, etc.), and $\Sigma(k)$ is the self-energy matrix:

$$\Sigma(k) = i \cdot N \cdot \text{Im}(R(k)) \tag{10.1}$$

$$R(k) = \frac{1}{N_{k'}} \sum_{k' \in BZ} T^\dagger(k, k') G^0_{k'} T(k', k), \tag{10.2}$$

where the real part of the self-energy matrix in Eq. (10.1) has been omitted, as it can be absorbed into the Fermi energy. The conductivity contributed by the bulk (surface) states, $\sigma_b$ ($\sigma_s$), is derived from Eq. (7) by integrating over the bulk-state (surface-state) $k$-points. Note that the impurity vertex corrections are neglected in Eq. (7).

## APPENDIX G: SIMULATION OF RESISTIVITY VS. THICKNESS SCALING

We normalize the slab resistivity ($\rho$) against the 3D resistivity of an infinite bulk without vacancy ($\rho_0$) to obtain the relative resistivity,

$$\frac{\rho}{\rho_0} = \frac{\sigma_0}{\left(\frac{\sigma^{2D}}{d}\right)}. \tag{11}$$

Here $\sigma^{2D}$ is the 2D conductivity of the slab model from Eq. (7); $d$ is the thickness of the slab; $\sigma_0$ is the 3D conductivity of the infinite bulk system and can be rewritten as:

$$\sigma_0 = 2\frac{e^2}{h} \int_{\vec{k}} \frac{d^3k}{(2\pi)^3} \text{Re Tr} \left[ \frac{\partial H}{\partial k_x} \left( G^a(\vec{k}, E_F) - G^r(\vec{k}, E_F) \right) \frac{\partial H}{\partial k_x} G^r(\vec{k}, E_F) \right] \tag{12}$$

where $H_{\vec{k}}$ is the TB Hamiltonian of bulk CoSi.

## APPENDIX H: BULK MEAN-FREE PATH CALCULATION

We calculate the near-equilibrium carrier transport to estimate the bulk mean-free path ($l_0$). First, we obtain the ballistic conductance $G_B$ of the bulk CoSi system defined as

$$G_B = G_0 \int dE\, T(E)\, \delta(E - E_F), \tag{13}$$

where $G_0$ is the conductance quantum and

$$T(E) = \int dk_y dk_z \, T(k_y, k_z, E) \tag{14}$$

is the total transmission of the pristine bulk CoSi. In Eq. (13), the $\delta$ function is broadened to a Lorentzian of width 1 meV, to be consistent with the 3D infinite bulk conductivity ($\sigma_0$) obtained by Eq. (12). Here, the $k$-resolved transmission $T(k_y, k_z, E)$ is calculated using the NEFG method (as discussed above for Eq. (3)) with a 300×300 k-grid-mesh over the Brillouin zone. Next, following the phenomenological expression

$$\sigma_0 = \sigma^B \frac{l_0}{a} \tag{15}$$

introduced in Ref. [48], where $\sigma^B = G_B a/A$ and $A$ is the cross-sectional area of the unit cell used to calculate $G_B$, we can obtain the bulk mean-free path as follows for fitting to the Fuchs-Sondheimer model in Fig. 4(b):

$$l_0 = \frac{\sigma_0 A}{G_B}. \tag{16}$$

**APPENDIX I: FITTING TO FUCHS-SONDHEIMER MODEL**

According to Fuchs-Sondheimer's theory[30], the ratio of the bulk resistivity ($1/\sigma_b$) of the film to that of the infinite bulk ($1/\sigma_0$) can be written as

$$\frac{\sigma_0}{\sigma_b(\kappa)} = \left(1 - \frac{3}{2\kappa}(1-p)\int_1^\infty \left(\frac{1}{t^3} - \frac{1}{t^5}\right)\frac{1-e^{-\kappa t}}{1-pe^{-\kappa t}}dt\right)^{-1}, \tag{17}$$

where $\kappa = d/l_0$, $l_0$ is the bulk mean free path calculated from Eq. (16), $d$ is thickness of the film, $p$ is the degree of specular reflection of the bulk electrons at the surface ($0 < p \leq 1$). Given $d$ and $l_0$, we can obtain $p$ by fitting to $\frac{\sigma_0}{\sigma_b(\kappa)}$.

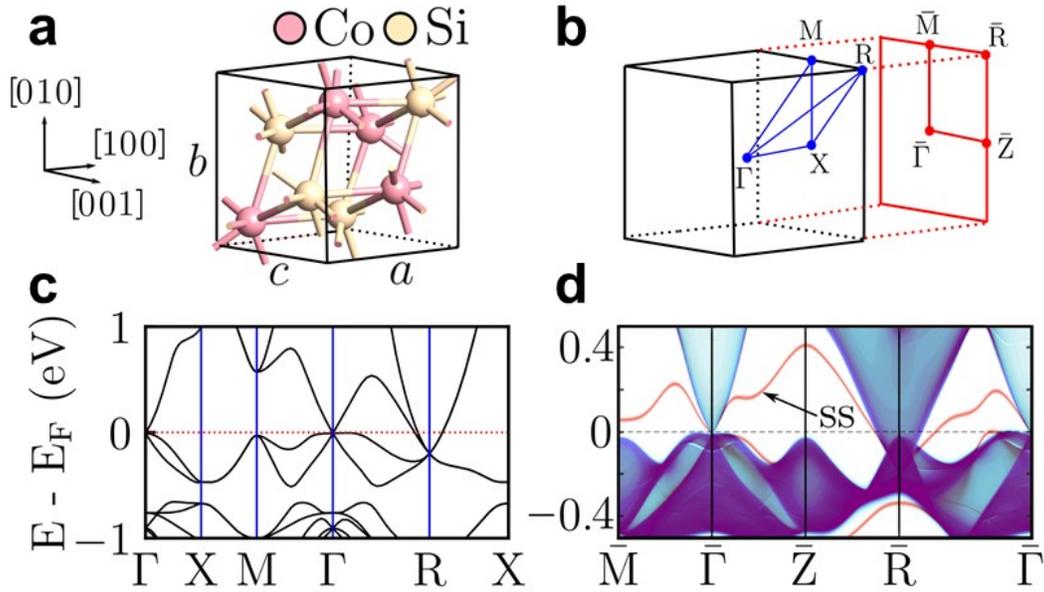

FIG. 1. Crystal structure and band structure of bulk and semi-infinite CoSi. (a) Unit cell of CoSi with lattice constants $a = b = c = 4.438$ Å. (b) Corresponding bulk Brillouin Zone (BZ) and its projection along the (100) surface (shown in red). (c) Bulk electronic band structure of CoSi. d, Spectral weight of a (100)-oriented semi-infinite CoSi slab, where the surface states (SS) are indicated in red.

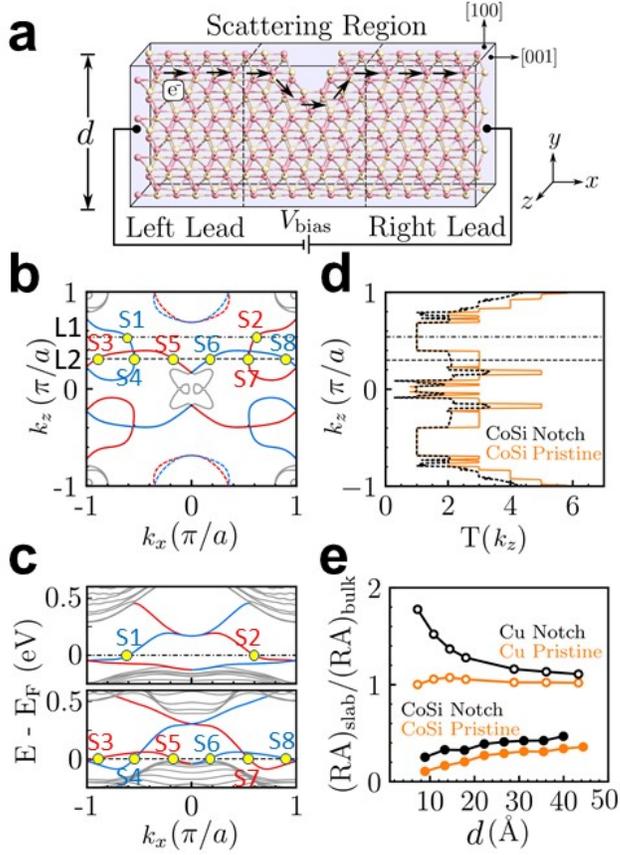

FIG. 2. First-principles informed quantum transport in CoSi slabs. (a) Schematic view of a two-terminal device with a scattering region composed of a CoSi slab of thickness $d$ along the [100] direction, sandwiched between two semi-infinite CoSi leads. An external bias-voltage $V_{bias} \ll E_F/e$ (with $e$ the electron's charge) injects a charge current in the [001] direction. The arrows illustrate the surface current distribution on the top surface in the presence of defects (shown as a notch). (b) Fermi surface of a 40AL CoSi slab with $d$ = 36.05 Å. Fermi-arc surface states on the top and bottom surfaces are denoted in red and blue solid lines, respectively. The dashed lines indicate the topologically trivial surface states, while the gray lines indicate the bulk states. (c) Top panel: band structure of the same slab along the linecut L1 in b at $k_z = 0.52\ (\pi/a) \equiv k_{z1}$, revealing the Fermi-arc surface states (in blue and red) connecting between the bulk valence and conduction bands (in gray). The markers $S_1$ and $S_2$ in b and c indicate the conducting channels available for transport at a fixed $k_{z1}$. Bottom panel: band structure of the slab along the linecut L2 in b at $k_z = 0.52\ (\pi/a) \equiv k_{z2}$. The markers $S_i$ ($i$ = 3 to 8) indicate the conducting channels available for transport at $k_{z2}$. (d) $k_z$-resolved transmission $T(k_z)$ for a pristine 40 AL CoSi slab, compared to that for a slab with a notch on the top surface. (e) Thickness dependence of the resistance-area (RA) product of Cu and CoSi slabs, normalized by the RA product in the infinite thickness limit, denoted as (RA)$_{bulk}$. The orange curves represent pristine slabs, while the black curves represent slabs with surface disorder in the form of a notch on the top surface.

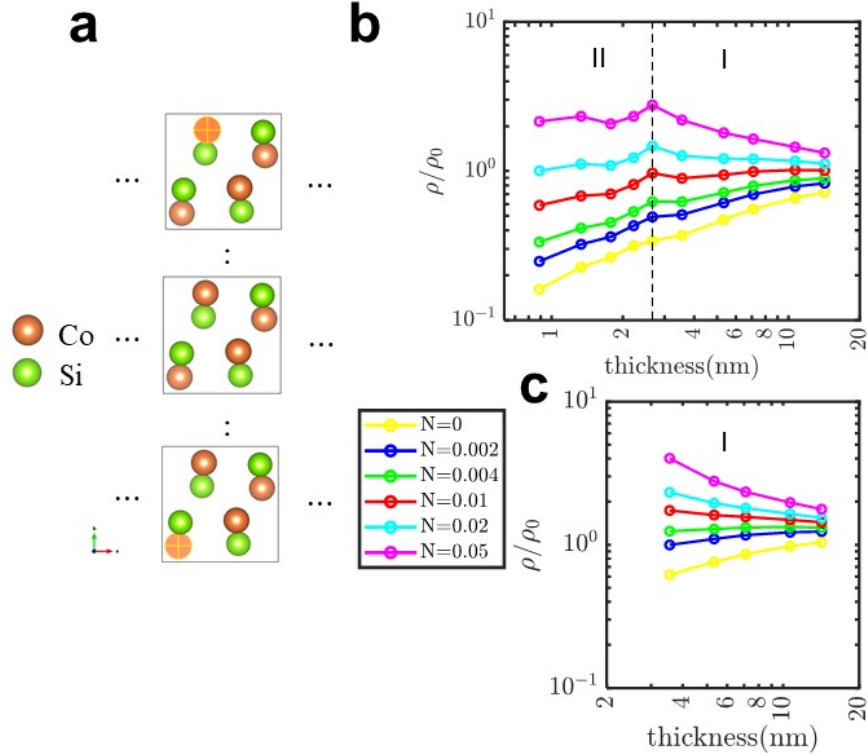

FIG. 3. Resistivity scaling in CoSi slabs with point defects. (a) Crystal structure of a CoSi slab. A yellow cross indicates the position of a surface vacancy point defect. (b) Longitudinal resistivity $\rho$ as a function of the slab thickness, calculated from first-principles electronic structure and Kubo's formula, for different surface defect density $N$ ($N = 0.01$ is equivalent to areal density $\sim 5 \times 10^{12}$ cm$^{-2}$). In region $I$, well-differentiated surface and bulk states coexist at the Fermi level. In region $II$, only remnants of the surface states are present at the Fermi level. The resistivity in the infinite thickness limit is denoted as $\rho_0$. (c) Longitudinal resistivity $\rho$ as a function of the slab thickness in region $I$, calculated from a combined analytical and first-principles approach (Supplemental Material Sec. S1), for different surface defect density $N$.

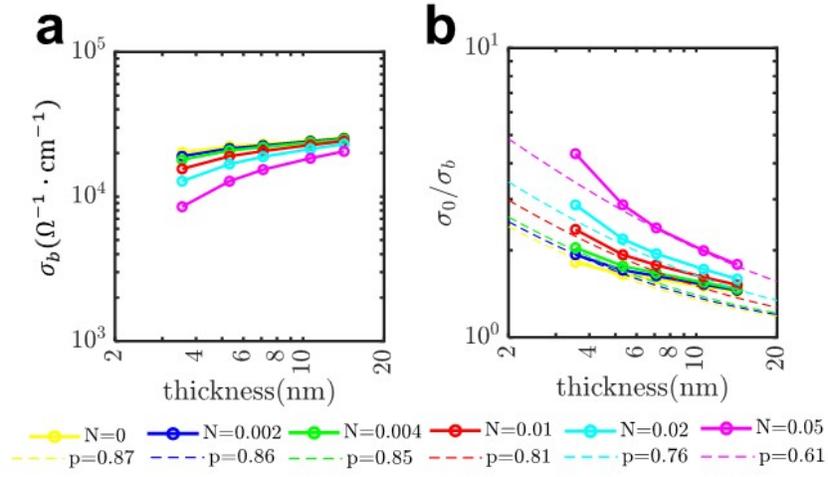

FIG. 4. Scaling of bulk-state conductivity in CoSi slabs. (a) Pure bulk conductivity $\sigma_b$ (not including the surface state contribution) as a function of slab thickness, for different surface defect density $N$ ($N = 0.01$ is equivalent to areal density $\sim 5 \times 10^{12}$ cm$^{-2}$). The range of thickness considered corresponds to region $I$ of Fig. 3. (b) Inverse ratio of the bulk conductivity $\sigma_b$ to the infinite thickness conductivity $\sigma_0$. The numerical data are fit to the Fuchs-Sondheimer model. The probability of specular surface scattering, $p$, decreases as $N$ increases.

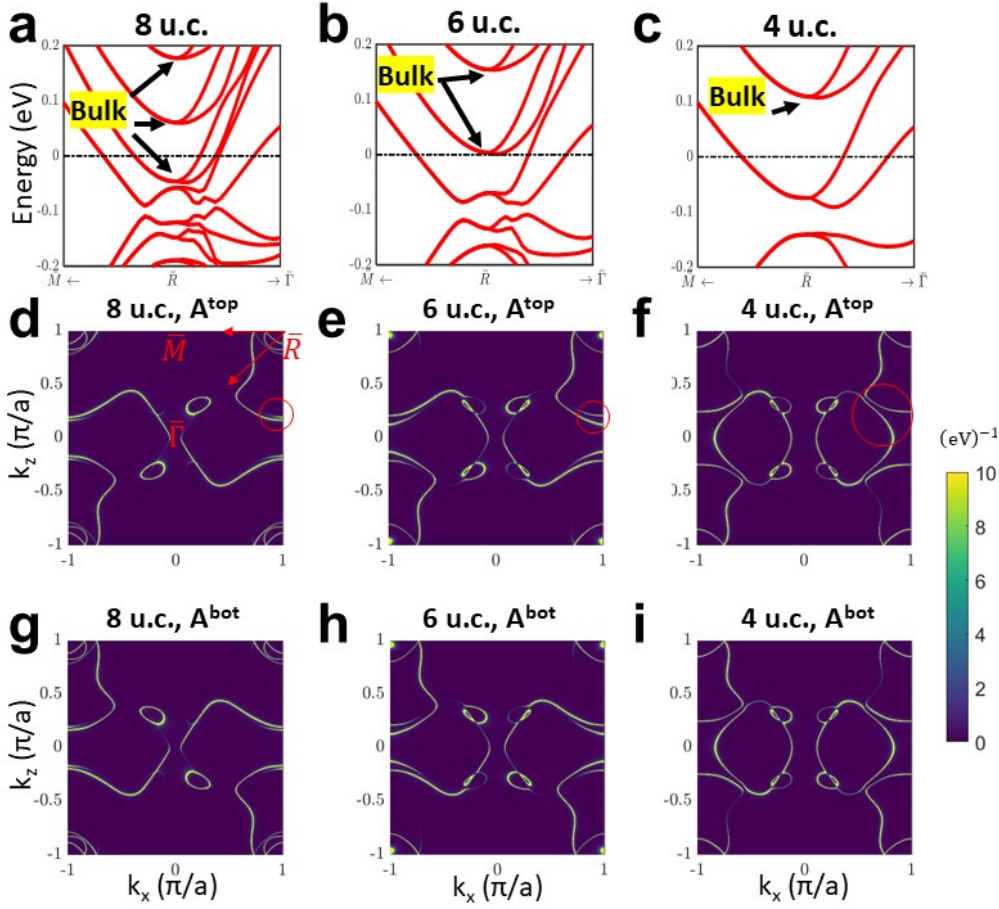

FIG. 5. Band structures and surface-resolved spectral weight of CoSi thin slabs. (a)-(c) Band structures near $\bar{R}$, for slab thicknesses of 4 unit cells (u. c.), 6 u. c. and 8 u. c. (respectively). The arrows show the evolution of the bottom of the bulk sub-bands near the Fermi level with decreasing thickness. At a thickness of 6 u. c., the bottom of a bulk sub-band crosses the Fermi level, giving rise to a van Hove singularity in the density of states at the Fermi level. Below 6 u. c., the only bands left at the Fermi level are the remnants of the surface states. (d)-(i) Electronic spectral functions of the top 2 layers ($A_k^{top}$) and bottom 2 layers ($A_k^{bot}$) in CoSi slabs with thicknesses of 4 u. c., 6 u. c., and 8 u. c. The two spectral functions are related by a $C_2$ rotation. In the thinnest samples, surface states are spread rather uniformly over the entire film volume, thereby yielding $A_k^{top} \simeq A_k^{bot}$. The red circles in (d)-(f) indicate the splitting of Fermi arcs due to top and bottom surface hybridization. The thinner the slab, the larger the splitting.

# Supplemental Material:
# Unconventional Resistivity Scaling in Topological Semimetal CoSi

Shang-Wei Lien†,[1] Ion Garate†*,[2] Utkarsh Bajpai,[3] Cheng-Yi Huang,[4] Chuang-Han Hsu,[5] Yi-Hsin Tu,[1] Nicholas A. Lanzillo,[3] Arun Bansil,[4] Tay-Rong Chang,[1,6,7] Gengchiau Liang*,[8] Hsin Lin*,[5] and Ching-Tzu Chen*[9]

[1] *Department of Physics, National Cheng Kung University, Tainan 701, Taiwan.*
[2] *Département de physique, Institut quantique and Regroupement Québécois sur les Matériaux de Pointe, Université de Sherbrooke, Sherbrooke, Québec, Canada J1K 2R1.*
[3] *IBM Research, 257 Fuller Road, Albany, NY 12203, USA.*
[4] *Department of Physics, Northeastern University, Boston, Massachusetts 02115, USA.*
[5] *Institute of Physics, Academia Sinica, Nankang Taipei 11529, Taiwan, Republic of China.*
[6] *Center for Quantum Frontiers of Research and Technology (QFort), Tainan 701, Taiwan.*
[7] *Physics Division, National Center for Theoretical Sciences, Hsinchu, Taiwan.*
[8] *Department of Electrical and Computer Engineering, Faculty of Engineering, National University of Singapore, Singapore, Singapore.*
[9] *IBM T.J. Watson Research Center, 1101 Kitchawan Road, Yorktown Heights, NY 10598, USA.*

(Dated: September 12, 2022)

†these authors contributed equally
*corresponding authors.
Email: ion.garate@usherbrooke.ca; elelg@nus.edu.sg; nilnish@gate.sinica.edu.tw; cchen3@us.ibm.com

**Outline**

S1. Analytical theory of the resistivity scaling and scattering lengths in CoSi films

S2. Supplementary figures

### S1. Analytical theory of the resistivity scaling and scattering lengths in CoSi films

In this section, we begin by adapting the theory of Breitkreiz and Brouwer [PRL **123**, 066804 (2019)] to calculate the bulk and surface contributions to the three dimensional current density in CoSi films. The validity of this theory is limited to thicker films, where the surface and the bulk states are well differentiated.

Then, we go beyond the theory of Breitkreiz and Brouwer by providing microscopic (Fermi golden rule) expressions for the relaxation lengths. In the Methods, these expressions are computed using first-principles electronic structure methods. In addition, these expressions allow to compare the analytical theory of the resistivity scaling with the fully numerical results based on Kubo's formula.

#### A. Generalities

We consider a film with spatial dimensions $L_x \times d \times L_z$ (Fig. S1), where $L_x$ and $L_z$ are assumed to be much larger than the film thickness $d$.

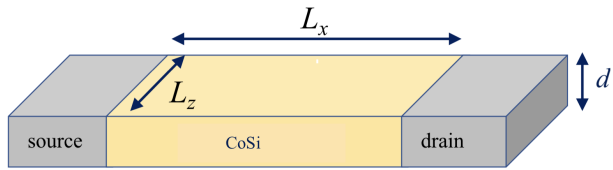

Figure S1: CoSi film connected to source and drain electrodes

In the bulk, CoSi hosts Fermi pockets around the $\Gamma$, $R$ and $M$ points of the three dimensional Brillouin zone. Unlike in the model considered by Breitkreiz and Brouwer, CoSi cannot be separated into two coupled subsystems that are related to one another by time-reversal. We will instead consider that the bulk is composed of three subsystems, with low-energy fermions near $\Gamma$, $R$ and $M$. Each subsystem is invariant under time-reversal (because $\Gamma$, $R$ and $M$ are time-reversal-invariant momenta).

At the surface, CoSi hosts Fermi arcs. There are two Fermi arcs on the $y = d/2$ surface, denoted as $s$ and $\bar{s}$, which connect the projections of the $\Gamma$ and $R$ points on the surface Brillouin zone. The group velocities of the two arcs are opposite to one another, due to time-reversal symmetry. For simplicity, we will assume these velocities to be constant (independent of momentum). Then, there are two more arcs at the $y = -d/2$ surface. These arcs have the opposite group velocities to the ones on the $y = d/2$ surface.

### B. Continuity equations

The current densities in the bulk (in A/m$^2$) can be written as

$$\mathbf{j}_\Gamma = eD_\Gamma n_\Gamma \boldsymbol{\nabla}\mu_\Gamma$$
$$\mathbf{j}_R = eD_R n_R \boldsymbol{\nabla}\mu_R$$
$$\mathbf{j}_M = eD_M n_M \boldsymbol{\nabla}\mu_M, \tag{S1}$$

where $D_\alpha$, $n_\alpha$ and $\mu_\alpha$ are the diffusion constants, the Fermi level density of states and the deviations of the electrochemical potentials from the equilibrium Fermi energy, respectively, at sector $\alpha = \Gamma, R, M$. We assume a diffusive transport in the bulk.

The surface current densities (in A/m) are

$$j_{s,\pm} = \pm e n_s v \mu_{s,\pm}$$
$$j_{\bar{s},\pm} = \mp e n_s v \mu_{\bar{s},\pm}, \tag{S2}$$

where $\pm$ stands for $y = \pm d/2$ surfaces, $n_s$ is the density of states of Fermi arcs at the Fermi energy, $v$ is the absolute value of the (constant) group velocity of surface electrons, and $\mu_{s\pm}$ is the deviation of the surface electrochemical potential from the equilibrium Fermi energy in the $y = \pm d/2$ surface. The surface states $s+$ and $\bar{s}+$ (or $s-$ and $\bar{s}-$) are related by time-reversal symmetry.

We assume that we apply a uniform electric field $E_x$ in the $x$ direction. We want to find out the expressions for the bulk and surface current in terms of $E_x$. This requires finding how all the $\mu-$s depend on $E_x$. To do so, we need to solve the following set of continuity equations (which are adapted from the theory of Breitkreiz and Brouwer):

$$\boldsymbol{\nabla}\cdot \mathbf{j}_\Gamma = e n_s v \left[\frac{\mu_\Gamma - \mu_{s,+}}{l_{s\Gamma}} + \frac{\mu_\Gamma - \mu_{\bar{s},+}}{l_{s\Gamma}}\right]\delta(y-d/2)$$
$$+ e n_s v \left[\frac{\mu_\Gamma - \mu_{s,-}}{l_{s\Gamma}} + \frac{\mu_\Gamma - \mu_{\bar{s},-}}{l_{s\Gamma}}\right]\delta(y+d/2) + e n_\Gamma v \left[\frac{\mu_\Gamma - \mu_R}{l_{\Gamma R}} + \frac{\mu_\Gamma - \mu_M}{l_{\Gamma M}}\right] \quad \text{(S3a)}$$

$$\boldsymbol{\nabla}\cdot \mathbf{j}_R = e n_s v \left[\frac{\mu_R - \mu_{s,+}}{l_{sR}} + \frac{\mu_R - \mu_{\bar{s},+}}{l_{sR}}\right]\delta(y-d/2)$$
$$+ e n_s v \left[\frac{\mu_R - \mu_{s,-}}{l_{sR}} + \frac{\mu_R - \mu_{\bar{s},-}}{l_{sR}}\right]\delta(y+d/2) + e n_\Gamma v \left[\frac{\mu_R - \mu_\Gamma}{l_{\Gamma R}} + \frac{\mu_R - \mu_M}{l_{RM}}\right] \quad \text{(S3b)}$$

$$\boldsymbol{\nabla}\cdot \mathbf{j}_M = e n_s v \left[\frac{\mu_M - \mu_{s,+}}{l_{sM}} + \frac{\mu_M - \mu_{\bar{s},+}}{l_{sM}}\right]\delta(y-d/2)$$
$$+ e n_s v \left[\frac{\mu_M - \mu_{s,-}}{l_{sM}} + \frac{\mu_M - \mu_{\bar{s},-}}{l_{sM}}\right]\delta(y+d/2) + e n_\Gamma v \left[\frac{\mu_M - \mu_\Gamma}{l_{\Gamma M}} + \frac{\mu_M - \mu_R}{l_{RM}}\right] \quad \text{(S3c)}$$

$$\partial_x j_{s,\pm} = e n_s v \left[\frac{\mu_{s,\pm} - \mu_\Gamma(\pm d/2)}{l_{s\Gamma}} + \frac{\mu_{s,\pm} - \mu_R(\pm d/2)}{l_{sR}} + \frac{\mu_{s,\pm} - \mu_M(\pm d/2)}{l_{sM}}\right]$$
$$+ e n_s v \frac{\mu_{s,\pm} - \mu_{\bar{s},\pm}}{l_{s\bar{s}}} + e n_s v \frac{\mu_{s,\pm} - \mu_{s,\mp}}{l_{\text{tb}}} + e n_s v \frac{\mu_{s,\pm} - \mu_{\bar{s},\mp}}{l_{\text{tb}}} \quad \text{(S3d)}$$

$$\partial_x j_{\bar{s},\pm} = e n_s v \left[\frac{\mu_{\bar{s},\pm} - \mu_\Gamma(\pm d/2)}{l_{s\Gamma}} + \frac{\mu_{\bar{s},\pm} - \mu_R(\pm d/2)}{l_{sR}} + \frac{\mu_{\bar{s},\pm} - \mu_M(\pm d/2)}{l_{sM}}\right]$$
$$+ e n_s v \frac{\mu_{\bar{s},\pm} - \mu_{s,\pm}}{l_{s\bar{s}}} + e n_s v \frac{\mu_{\bar{s},\pm} - \mu_{\bar{s},\mp}}{l_{\text{tb}}} + e n_s v \frac{\mu_{\bar{s},\pm} - \mu_{s,\mp}}{l_{\text{tb}}}. \quad \text{(S3e)}$$

Equations (S3a), (S3b) and (S3c) are the continuity equations for bulk electrons at $\Gamma$, $R$ and $M$, respectively, in the steady state. The terms in the right hand side describe processes that break the conservation of the number of particles in sector $\alpha$. For example, for the charge near $\Gamma$, an electron can jump onto one of the two surfaces (this is possible only in the immediate vicinity of the surfaces, as otherwise the matrix element for the transition vanishes), or it can jump either to $R$ or to $M$ Fermi pockets. The net rate of these jumps is proportional to the difference in electrochemical potentials between the different subsystems; this fact will be justified below from microscopic theory via the Fermi golden rule. The constant of proportionality contains the phenomenological relaxation lengths. We note that $\boldsymbol{\nabla}\cdot \mathbf{j}_\Gamma + \boldsymbol{\nabla}\cdot \mathbf{j}_R + \boldsymbol{\nabla}\cdot \mathbf{j}_M = 0$ for all $y \neq \pm d/2$. This simply states that the total bulk charge must be conserved in the absence of surfaces.

Equations (S3d) and (S3e) describe the continuity equations for the surface electrons, in the steady state. Once again, the surface charge is not conserved due to the couplings between the Fermi arcs and the bulk sectors $\alpha$. We have also incorporated direct coupling between the top and the bottom surface through the relaxation length $l_{\text{tb}}$. Time-reversal symmetry requires that $l_{\bar{s}\alpha} = l_{s\alpha}$ (recall that $\Gamma$, $R$ and $M$ are invariant under time-reversal, whereas $s$ goes to $\bar{s}$ under time-reversal). The notation $\mu_\alpha(\pm d/2)$ stands for $\mu_\alpha$ evaluated at $y = \pm d/2$.

## C. Solution of continuity equations

We seek a solution of the form

$$\mu_\Gamma = eE_x x + \tilde{\mu}_\Gamma(y) \tag{S4a}$$

$$\mu_R = eE_x x + \tilde{\mu}_R(y) \tag{S4b}$$

$$\mu_M = eE_x x + \tilde{\mu}_M(y) \tag{S4c}$$

$$\mu_{s,\pm} = eE_x x + \tilde{\mu}_{s,\pm}(y) \tag{S4d}$$

$$\mu_{\bar{s},\pm} = eE_x x + \tilde{\mu}_{\bar{s},\pm}(y), \tag{S4e}$$

where the $\tilde{\mu}$ are unknown functions of only $y$. The fact that $\mu_\Gamma$, $\mu_R$ and $\mu_M$ contain $eE_x x$ makes sense because the bulk transport is assumed to be diffusive. The fact that $\mu_{s,\pm}$ and $\mu_{\bar{s},\pm}$ contains the same factor also makes sense since Fermi arcs are coupled to bulk states (it is assumed that the bulk-surface scattering lengths are shorter than the channel length $L_x$).

We begin by determining the bulk electrochemical potentials. Combining Eqs. (S1) and (S4), we have

$$j_{\Gamma,x} = e^2 D_\Gamma n_\Gamma E_x \tag{S5a}$$

$$j_{R,x} = e^2 D_R n_R E_x \tag{S5b}$$

$$j_{M,x} = e^2 D_M n_M E_x \tag{S5c}$$

$$j_{\Gamma,y} = e D_\Gamma n_\Gamma \partial_y \tilde{\mu}_\Gamma \tag{S5d}$$

$$j_{R,y} = e D_R n_R \partial_y \tilde{\mu}_R \tag{S5e}$$

$$j_{M,y} = e D_M n_M \partial_y \tilde{\mu}_M. \tag{S5f}$$

For all $y \neq \pm d/2$, Eqs. (S3a), (S3b) and (S3c) give

$$\partial_y j_{\Gamma,y} = e n_\Gamma v \left[ \frac{\tilde{\mu}_\Gamma - \tilde{\mu}_R}{l_{\Gamma R}} + \frac{\tilde{\mu}_\Gamma - \tilde{\mu}_M}{l_{\Gamma M}} \right]$$

$$\partial_y j_{R,y} = e n_\Gamma v \left[ \frac{\tilde{\mu}_R - \tilde{\mu}_\Gamma}{l_{\Gamma R}} + \frac{\tilde{\mu}_R - \tilde{\mu}_M}{l_{RM}} \right]$$

$$\partial_y j_{M,y} = e n_\Gamma v \left[ \frac{\tilde{\mu}_M - \tilde{\mu}_\Gamma}{l_{\Gamma M}} + \frac{\tilde{\mu}_M - \tilde{\mu}_R}{l_{RM}} \right], \tag{S6}$$

where we used $\partial_x j_{\Gamma,x} = \partial_x j_{R,x} = \partial_x j_{M,x} = 0$ (because $E_x$ is uniform). Combining Eqs. (S5) and (S6), we have

$$\partial_y^2 \tilde{\mu}_\Gamma = \frac{1}{\lambda_{\Gamma R}^2} (\tilde{\mu}_\Gamma - \tilde{\mu}_R) + \frac{1}{\lambda_{\Gamma M}^2} (\tilde{\mu}_\Gamma - \tilde{\mu}_M)$$

$$\partial_y^2 \tilde{\mu}_R = \frac{1}{\lambda_{R\Gamma}^2} (\tilde{\mu}_R - \tilde{\mu}_\Gamma) + \frac{1}{\lambda_{RM}^2} (\tilde{\mu}_R - \tilde{\mu}_M)$$

$$\partial_y^2 \tilde{\mu}_M = \frac{1}{\lambda_{M\Gamma}^2} (\tilde{\mu}_M - \tilde{\mu}_\Gamma) + \frac{1}{\lambda_{MR}^2} (\tilde{\mu}_M - \tilde{\mu}_R), \tag{S7}$$

where we have defined

$$\lambda_{\Gamma R} = \sqrt{\frac{D_\Gamma l_{\Gamma R}}{v}}$$

$$\lambda_{\Gamma M} = \sqrt{\frac{D_\Gamma l_{\Gamma M}}{v}}$$

$$\lambda_{R\Gamma} = \sqrt{\frac{n_R D_R l_{\Gamma R}}{v n_\Gamma}}$$

$$\lambda_{RM} = \sqrt{\frac{n_R D_R l_{RM}}{v n_\Gamma}}$$

$$\lambda_{M\Gamma} = \sqrt{\frac{n_M D_M l_{\Gamma M}}{v n_\Gamma}}$$

$$\lambda_{MR} = \sqrt{\frac{n_M D_M l_{RM}}{v n_\Gamma}}. \tag{S8}$$

The general solution of Eq. (S7) is cumbersome; we will not reproduce it here. Instead, we will impose constraints based on time-reversal symmetry, which greatly simplify the solution. In a time-reversal invariant system placed under a uniform electric field that points along a high-symmetry direction of the crystal, there cannot be any net transverse current density (at least to linear order in the electric field). Moreover, since $\Gamma$, $R$ and $M$ are time-reversal-invariant momenta, time-reversal symmetry implies $j_{\Gamma,y}(y) = -j_{\Gamma,y}(y)$, $j_{R,y}(y) = -j_{R,y}(y)$ and $j_{M,y}(y) = -j_{M,y}(y)$. Accordingly, $j_{\Gamma,y}(y) = j_{R,y}(y) = j_{M,y}(y) = 0$ for all $y$, and hence

$$\tilde{\mu}_\Gamma = \tilde{\mu}_R = \tilde{\mu}_M = \mu_0 = \text{const.} \tag{S9}$$

The fact that the three bulk electrochemical potentials are identical to one another follows from the solution of Eq. (S7) under the condition $\partial_y \mu_\alpha = 0$ for all $\alpha$. It also makes sense from the fact that there is no chiral anomaly in the absence of an external magnetic field. Below, we will see that the value of $\mu_0$ does not matter.

Next, we concentrate on the surface electrochemical potentials. Combining Eqs. (S13), (S3d), (S3e), (S4) and (S9), we get

$$\pm eE_x = (\tilde{\mu}_{s,\pm} - \mu_0)\left(\frac{1}{l_{s\Gamma}} + \frac{1}{l_{sR}} + \frac{1}{l_{sM}}\right) + \frac{\tilde{\mu}_{s,\pm} - \tilde{\mu}_{\bar{s},\pm}}{l_{s\bar{s}}} + \frac{\mu_{s,\pm} - \mu_{s,\mp}}{l_{tb}} + \frac{\mu_{s,\pm} - \mu_{\bar{s},\mp}}{l_{tb}}$$

$$\mp eE_x = (\tilde{\mu}_{\bar{s},\pm} - \mu_0)\left(\frac{1}{l_{s\Gamma}} + \frac{1}{l_{sR}} + \frac{1}{l_{sM}}\right) + \frac{\tilde{\mu}_{\bar{s},\pm} - \tilde{\mu}_{s,\pm}}{l_{s\bar{s}}} + e\frac{\mu_{\bar{s},\pm} - \mu_{\bar{s},\mp}}{l_{tb}} + \frac{\mu_{\bar{s},\pm} - \mu_{s,\mp}}{l_{tb}}. \tag{S10}$$

Solving these equations, we get

$$\tilde{\mu}_{s,\pm} = \mu_0 \pm \frac{eE_x}{\frac{1}{l_{s\Gamma}} + \frac{1}{l_{sR}} + \frac{1}{l_{sM}} + \frac{2}{l_{s\bar{s}}} + \frac{2}{l_{tb}}}$$

$$\tilde{\mu}_{\bar{s},\pm} = \mu_0 \mp \frac{eE_x}{\frac{1}{l_{s\Gamma}} + \frac{1}{l_{sR}} + \frac{1}{l_{sM}} + \frac{2}{l_{s\bar{s}}} + \frac{2}{l_{tb}}}. \tag{S11}$$

Equations (S4), (S9) and (S11) provide the complete solution for the electrochemical potentials. In order to check the consistency of the solutions, let us integrate Eq. (S3a) in the immediate vicinity of the $y = d/2$ surface, giving

$$j_{\Gamma,y}(d/2+0^+) - j_{\Gamma,y}(d/2-0^+) = en_s v \frac{\mu_\Gamma(d/2) - \mu_{s,+}}{l_{s\Gamma}} + en_s v \frac{\mu_\Gamma(d/2) - \mu_{\bar{s},+}}{l_{s\Gamma}}, \tag{S12}$$

where $0^+$ is an infinitesimal positive number. In view of our solutions for the chemical potentials, Eq. (S12) simply yields $0 = 0$. The same conclusion applies if we integrate Eq. (S3a) in the immediate vicinity of the $y = -d/2$ surface, or if we integrate either Eq. (S3b) or Eq. (S3c) in the immediate vicinity of either surface. In the calculation of Breitkreiz and Brouwer, this integration was necessary in order to fully determine the electrochemical potentials. In our present case, we have already determined them completely without having to do any boundary integral. Indeed, this boundary integral gives no new information in our case.

### D. Surface vs bulk contributions to the average current density

We are now ready to write the final expressions for the currents. For example, the average contribution from Fermi arcs to the three dimensional current density is

$$j_s = \frac{j_{s,+} + j_{s,-} + j_{\bar{s},+} + j_{\bar{s},-}}{d} = \frac{4e^2 n_s v E_x}{d\left(\frac{1}{l_{s\Gamma}} + \frac{1}{l_{sR}} + \frac{1}{l_{sM}} + \frac{2}{l_{s\bar{s}}} + \frac{2}{l_{\text{tb}}}\right)}. \tag{S13}$$

For the range of film thicknesses that we consider, the direct surface-to-surface tunneling is weak and therefore $1/l_{\text{tb}}$ can be neglected in Eq. (S13). Moreover, using $n_s v = k_0/(2\pi h)$, where $k_0$ is the length of a straight line connecting the two ends of a Fermi arc in momentum space (see the Supplemental Material in the article of Breitkreiz and Brouwer), Eq. (S13) can be rewritten as

$$j_s = \frac{2}{\pi h} e^2 k_0 \frac{l_s}{d} E_x, \tag{S14}$$

where we have defined an effective surface scattering length $l_s$ via

$$\frac{1}{l_s} = \frac{1}{l_{s\Gamma}} + \frac{1}{l_{sR}} + \frac{1}{l_{sM}} + \frac{2}{l_{s\bar{s}}}. \tag{S15}$$

In our case, $k_0 = \sqrt{2}\pi/a$, where $a$ is the lattice constant of CoSi. We have quoted Eq. (S14) in the discussion of Fig. 3 of the main text.

Thus, we find that in CoSi the contribution from Fermi arcs to the average three dimensional current density will be inversely proportional to the film thickness. This result matches qualitatively with that of Breitkreiz and Brouwer's result if we take $l_{b\bar{b}} \ll W$ in their theory ($l_{b\bar{b}}$ is the scattering length between the time-reversed partner bulk states). This perhaps not surprising, if one thinks of CoSi as two "glued" time-reversed partner subsystems.

The dimensionless ratio between the bulk and the (average) surface current densities reads

$$\frac{j_s}{j_b} = \frac{\sigma_s}{\sigma_b} = \frac{2}{\pi h} e^2 k_0 \frac{l_s}{d} \frac{1}{\sigma_b}, \tag{S16}$$

where

$$\sigma_b = D_\Gamma n_\Gamma + D_R n_R + D_M n_M \tag{S17}$$

is the conductivity of the bulk states.

The ratio $\sigma_s/\sigma_b$ plays an important role in the interpretation of the Fig. 3 of the main text. For sufficiently small $d$, the surface current density is dominant and $\sigma_s > \sigma_b$. Conversely, as $d$ gets larger, the relative importance of the bulk current density grows and $\sigma_s < \sigma_b$. To be more precise, in the theory of Fuchs and Sondheimer (see main text), the bulk conductivity scales as

$$\sigma_b \sim \sigma_0 \frac{1+p}{1-p} \frac{d}{l_0} \ln\left(\frac{l_0}{d}\right), \tag{S18}$$

where $l_0$ is the bulk mean free path, $p < 1$ is the probability of specular scattering at the surface, and $\sigma_0$ is the conductivity of an infinitely thick crystal ($\sigma_0 > \sigma_b$). This approximate expression holds for $l_0 \gg d$, which is the regime of interest in our case (see Supplementary Figure 4a). We may estimate $\sigma_0$ by assuming that the main contribution to it comes from the Fermi pocket enclosing the $R$ point (where the number of carriers is highest). Then, a simple estimate gives

$$\sigma_0 \sim \frac{e^2}{h} k_F^2 l_0, \tag{S19}$$

where $k_F$ is the Fermi wave vector of the bulk crystal for the Fermi pocket centered at $R$ (which for the purposes of the estimate we take to be spherical). Accordingly,

$$\frac{\sigma_s}{\sigma_b} \sim \frac{1-p}{1+p} \frac{k_0 l_s}{(k_F d)^2 \ln\left(\frac{l_0}{d}\right)}. \tag{S20}$$

Thus, the main factors enhancing the importance of the surface contribution to the average 3D conductivity are (i) long Fermi arcs ($k_0 \gg k_F$), (ii) long surface scattering lengths ($l_s \gg d$) and (iii) thin films ($d \lesssim k_F^{-1}$). When the surface defect density $N$ increases, $p$, $l_0$ and $l_s$ all decrease, but it is the decrease of $l_s$ that dominates the behavior of $\sigma_s/\sigma_b$ as a function of $N$ (see Supplementary Figures 3, 4 and 5).

### E. Fermi golden rule expressions for the relaxation lengths

Let us determine from microscopic theory the phenomenological relaxation lengths ($l_{s\bar{s}}$, $l_{s\Gamma}$, $l_{sR}$ and $l_{sM}$) appearing in the preceding subsection. For concreteness, we will begin with the relaxation length $l_{s\Gamma}$ associated to the coupling between the bulk valley $\Gamma$ and the surface band $s+$.

The *net* rate of charge transfer (in coulomb per second) from $\Gamma$ to $s+$, due to static disorder, can be approximated by the Fermi's golden rule expression

$$\left(\frac{\partial Q}{\partial t}\right)_{\Gamma \to s+} - \left(\frac{\partial Q}{\partial t}\right)_{s+ \to \Gamma} = e \frac{2\pi}{\hbar} \sum_{n \in \Gamma} \sum_{n' \in s+} \sum_{\mathbf{k}} \sum_{\mathbf{k'}} |\langle n, \mathbf{k}|V|n', \mathbf{k'}\rangle|^2 \left(f_{n,\mathbf{k}} - f_{n',\mathbf{k'}}\right) \delta\left(\epsilon_{n,\mathbf{k}} - \epsilon_{n',\mathbf{k'}}\right), \tag{S21}$$

where $n$ and $n'$ are the band labels for the $\Gamma$ and $s+$ electrons (respectively), $|n, \mathbf{k}\rangle$ and $|n', \mathbf{k}'\rangle$ are the corresponding eigenstates (note that $\mathbf{k}$ and $\mathbf{k}'$ are 2D momenta in the $xz$ plane), $V$ is the impurity potential, and $f_{n,\mathbf{k}} - f_{n',\mathbf{k}'}$ is the difference in electron occupation between states $(n, \mathbf{k})$ and $(n', \mathbf{k}')$. Assuming that the deviations from the equilibrium chemical potential are small, a Taylor expansion gives

$$\delta\left(\epsilon_{n,\mathbf{k}} - \epsilon_{n',\mathbf{k}'}\right)\left(f_{n,\mathbf{k}} - f_{n,\mathbf{k}'}\right) \simeq \delta\left(\epsilon_{n,\mathbf{k}} - \epsilon_{n',\mathbf{k}'}\right)(\mu_{s+} - \mu_\Gamma)\frac{\partial}{\partial\epsilon_{n,\mathbf{k}}}\frac{1}{e^{\beta(\epsilon_{n,\mathbf{k}} - \epsilon_F)} + 1}$$

$$\simeq (\mu_\Gamma - \mu_{s+})\delta\left(\epsilon_{n,\mathbf{k}} - \epsilon_{n',\mathbf{k}'}\right)\delta\left(\epsilon_{n,\mathbf{k}} - \epsilon_F\right)$$

$$= (\mu_\Gamma - \mu_{s+})\delta\left(\epsilon_{n',\mathbf{k}'} - \epsilon_F\right)\delta\left(\epsilon_{n,\mathbf{k}} - \epsilon_F\right), \tag{S22}$$

where $\epsilon_F$ is the Fermi energy in equilibrium and we have made a low-temperature approximation. Then, we have

$$\left(\frac{\partial Q}{\partial t}\right)_{\Gamma \to s+} - \left(\frac{\partial Q}{\partial t}\right)_{s+ \to \Gamma} = (\mu_\Gamma - \mu_{s+})e\frac{2\pi}{\hbar}\sum_{n\in\Gamma}\sum_{n'\in s+}\sum_{\mathbf{k}}\sum_{\mathbf{k}'}|\langle n, \mathbf{k}|V|n', \mathbf{k}'\rangle|^2 \delta\left(\epsilon_{n,\mathbf{k}} - \epsilon_F\right)\delta\left(\epsilon_{n',\mathbf{k}'} - \epsilon_F\right). \tag{S23}$$

On the other hand, the right hand side of the continuity equation Eq. (S3a) leads to

$$\left(\frac{\partial Q}{\partial t}\right)_{\Gamma \to s+} - \left(\frac{\partial Q}{\partial t}\right)_{s+ \to \Gamma} = \int d^3 r \left[\left(\frac{\partial \rho}{\partial t}\right)_{\Gamma \to s+} - \left(\frac{\partial \rho}{\partial t}\right)_{s+ \to \Gamma}\right] = \int d^3 r\, e n_s v \frac{\mu_\Gamma - \mu_{s,+}}{l_{s\Gamma}}\delta(y - d/2) = L_z L_x e n_s v \frac{\mu_\Gamma - \mu_{s,+}}{l_{s\Gamma}}, \tag{S24}$$

where $\rho$ is the charge density (C/m$^3$) and we have used the fact that $\mu_\Gamma - \mu_{s,+}$ is independent of $x$ and $z$ (note that both $\mu_\Gamma$ and $\mu_{s,+}$ depend on $x$, but their difference does not).

Identifying Eq. (S23) with Eq. (S24), we conclude that

$$\frac{e n_s v}{l_{s\Gamma}} = \frac{1}{L_x L_z}e\frac{2\pi}{\hbar}\sum_{n\in\Gamma}\sum_{n'\in s+}\sum_{\mathbf{k}}\sum_{\mathbf{k}'}|\langle n, \mathbf{k}|V|n', \mathbf{k}'\rangle|^2 \delta\left(\epsilon_{n,\mathbf{k}} - \epsilon_F\right)\delta\left(\epsilon_{n',\mathbf{k}'} - \epsilon_F\right). \tag{S25}$$

The scattering lengths $l_{sR}$ and $l_{sM}$ obey the same expression, upon replacing $\Gamma$ by $R$ and $M$ (respectively). Similarly, we obtain

$$\frac{e n_s v}{l_{s\bar{s}}} = \frac{1}{L_x L_z}e\frac{2\pi}{\hbar}\sum_{n\in s}\sum_{n'\in\bar{s}}\sum_{\mathbf{k}}\sum_{\mathbf{k}'}|\langle n, \mathbf{k}|V|n', \mathbf{k}'\rangle|^2 \delta\left(\epsilon_{n,\mathbf{k}} - \epsilon_F\right)\delta\left(\epsilon_{n',\mathbf{k}'} - \epsilon_F\right). \tag{S26}$$

For completeness, let us also discuss the bulk-to-bulk scattering lengths (such as $l_{\Gamma R}$), even though they are not needed for our theory. The counterpart of Eq. (S24) for these is

$$\left(\frac{\partial Q}{\partial t}\right)_{\Gamma \to R} - \left(\frac{\partial Q}{\partial t}\right)_{R \to \Gamma} = \int d^3 r \left[\left(\frac{\partial \rho}{\partial t}\right)_{\Gamma \to R} - \left(\frac{\partial \rho}{\partial t}\right)_{R \to \Gamma}\right] = \int d^3 r\, e n_\Gamma v \frac{\mu_\Gamma - \mu_R}{l_{\Gamma R}} = L_x L_z d e n_\Gamma v \frac{\mu_\Gamma - \mu_\Gamma}{l_{\Gamma R}}, \tag{S27}$$

which results in

$$\frac{e n_\Gamma v}{l_{\Gamma R}} = \frac{1}{L_x L_z d}e\frac{2\pi}{\hbar}\sum_{n\in\Gamma}\sum_{n'\in R}\sum_{\mathbf{k}}\sum_{\mathbf{k}'}|\langle n, \mathbf{k}|V|n', \mathbf{k}'\rangle|^2 \delta\left(\epsilon_{n,\mathbf{k}} - \epsilon_F\right)\delta\left(\epsilon_{n',\mathbf{k}'} - \epsilon_F\right). \tag{S28}$$

Note that the dimensions are correct: $n_s$ is the density of states per unit area of the surface states (an intensive quantity independent of system size unless the film is too thin) and $n_\Gamma$ is the density of states per unit volume of the bulk states (which is also an intensive quantity).

In the first principles calculations of the scattering lengths (see Methods) the impurity potential V is replaced by the T-matrix,

$$\begin{aligned}\frac{e n_s v}{l_{s\Gamma}} &= \frac{1}{L_x L_z} e \frac{2\pi}{\hbar} \sum_{n \in \Gamma} \sum_{n' \in s+} \sum_{\mathbf{k}} \sum_{\mathbf{k}'} |\langle n, \mathbf{k}|T|n', \mathbf{k}'\rangle|^2 \delta\left(\epsilon_{n,\mathbf{k}} - \epsilon_F\right) \delta\left(\epsilon_{n',\mathbf{k}'} - \epsilon_F\right) \\ \frac{e n_s v}{l_{s\bar{s}}} &= \frac{1}{L_x L_z} e \frac{2\pi}{\hbar} \sum_{n \in s} \sum_{n' \in \bar{s}} \sum_{\mathbf{k}} \sum_{\mathbf{k}'} |\langle n, \mathbf{k}|T|n', \mathbf{k}'\rangle|^2 \delta\left(\epsilon_{n,\mathbf{k}} - \epsilon_F\right) \delta\left(\epsilon_{n',\mathbf{k}'} - \epsilon_F\right).\end{aligned} \quad (S29)$$

together with $n_s v = k_0/(2\pi h)$ and $k_0 = \sqrt{2}\pi/a$. These expressions allow to go beyond the leading order in the impurity potential.

### F. Fermi golden rule expression for the bulk mean free path

Above, as well as in the main text, we make reference to the bulk mean free path $l_0$. In this section, we provide a Fermi golden rule expression for $l_0$, which will allow us in the next subsection to determine how $l_0$ scales with the system size and impurity density.

The starting point is the elastic scattering rate for an electron in a bulk state $|n, \mathbf{k}\rangle$:

$$\frac{1}{\tau_{n,\mathbf{k}}} = \frac{2\pi}{\hbar} \sum_{\mathbf{k}',n'} |\langle n, \mathbf{k}|V|n', \mathbf{k}'\rangle|^2 \delta(E_{\mathbf{k},n} - E_{\mathbf{k}',n'}), \quad (S30)$$

where $|n', \mathbf{k}\rangle$ are bulk states. Disregarding the distinction between transport and momentum scattering times (which is unimportant for our purposes below), the bulk mean free path in a cubic crystal can be approximated with a Fermi-surface average of $v^x_{n,\mathbf{k}} \tau_{n,\mathbf{k}}$,

$$l_0 \simeq \frac{\sum_{\mathbf{k},n} v^x_{n,\mathbf{k}} \tau_{n,\mathbf{k}} \delta(\epsilon_F - E_{n,\mathbf{k}})}{\sum_{\mathbf{k},n} \delta(\epsilon_F - E_{n,\mathbf{k}})}, \quad (S31)$$

where $v^x_{n,\mathbf{k}}$ is the $x-$component of the electronic group velocity in state $|n, \mathbf{k}\rangle$.

### G. Scaling of scattering lengths with impurity concentration and system size

To conclude this section, we analyze how the scattering lengths from the preceding subsection vary with the film thickness. This scaling is useful to understand the numerical results shown in Figs. 3 and 4 of the main text. For concreteness, let us see how the surface-to-bulk scattering length $l_{s\Gamma}$ scales with sample dimensions, when the latter are large. Later, we will adapt the arguments to other scattering lengths of interest.

First, we write the Bloch states as

$$|n, \mathbf{k}\rangle = \frac{1}{\sqrt{L_x L_z d}} e^{i\mathbf{k}\cdot\mathbf{r}} |u_{n,\mathbf{k}}\rangle, \quad (S32)$$

where $|u_{\mathbf{k},n}\rangle$ is the periodic part of the Bloch function (periodicity is present only in the $xz$ plane), with normalization

$$1 = \langle n,\mathbf{k}|n,\mathbf{k}\rangle = \frac{1}{L_x L_z d}\int_{\text{all}} d^3 r |u_{n,\mathbf{k}}(\mathbf{r})|^2 = \frac{N_{\text{cell}}}{L_x L_z d}\int_{\text{cell}} d^3 r |u_{n,\mathbf{k}}(\mathbf{r})|^2 = \frac{1}{V_{\text{cell}}}\int_{\text{cell}} d^3 r |u_{n,\mathbf{k}}(\mathbf{r})|^2, \quad (S33)$$

where $N_{\text{cell}}$ is the number of unit cells in the film and $V_{\text{cell}}$ is the volume of a unit cell. From this equation, we conclude that $u_{n,\mathbf{k}}(\mathbf{r})$ does not scale with $L_x$, $d$ and $L_z$ for a bulk state, but $u_{n',\mathbf{k}'}(\mathbf{r}) \sim \sqrt{d}$ for the surface state. The reason for this is that, for a surface state, $u_{n',\mathbf{k}'}(\mathbf{r})$ has a finite range along $y$. Thus, $|u_{n',\mathbf{k}'}(\mathbf{r})|^2 \sim d$ is needed to cancel the $1/d$ factor coming from $V_{\text{cell}}$ and to normalize the wave function to unity.

Second, we analyze the matrix element of the impurity potential. For a single short-range scatterer located at $\mathbf{r}_0$, $V(\mathbf{r}) = V_0 \delta(\mathbf{r}-\mathbf{r_0})$ and

$$\langle n,\mathbf{k}|V|n',\mathbf{k}'\rangle = \frac{1}{L_x L_z d}\int_{\text{all}} e^{i(\mathbf{k}'-\mathbf{k})\cdot\mathbf{r}}V_0 \delta(\mathbf{r}-\mathbf{r}_0)u_{n,\mathbf{k}}^*(\mathbf{r})u_{n',\mathbf{k}'}(\mathbf{r}) = \frac{1}{L_x L_z d}V_0 e^{i(\mathbf{k}'-\mathbf{k})\cdot\mathbf{r}_0}u_{n,\mathbf{k}}^*(\mathbf{r}_0)u_{n',\mathbf{k}'}(\mathbf{r}_0). \quad (S34)$$

Accordingly,

$$|\langle n,\mathbf{k}|V|n',\mathbf{k}'\rangle|^2 = \frac{V_0^2}{L_x^2 L_z^2 d^2}|u_{n,\mathbf{k}}(\mathbf{r}_0)|^2 |u_{n',\mathbf{k}'}(\mathbf{r}_0)|^2. \quad (S35)$$

Importantly, this quantity is proportional to the probability of finding a bulk electron and a surface electron simultaneously at $\mathbf{r}_0$ (recall that $n$ is a bulk state and $n'$ is a surface state, since we are analyzing $l_{s\Gamma}$). With system size, it scales as

$$|\langle n,\mathbf{k}|V|n',\mathbf{k}'\rangle|^2 \sim \frac{1}{L_x^2 L_z^2 d}. \quad (S36)$$

Third, consider $N_{\text{imp}}$ short-range impurities, $V(\mathbf{r}) = V_0 \sum_{i=1}^{N_{\text{imp}}} \delta(\mathbf{r}-\mathbf{r}_i)$. Assuming that these impurities are uncorrelated, the corresponding surface-to-bulk scattering rates can be added. Assuming that the impurities are randomly distributed in space with a density that is uniform on average, then only a fraction $N_{\text{imp}}r/d$ of them will contribute to $|\langle n,\mathbf{k}|V|n',\mathbf{k}'\rangle|^2$, where $r$ is the range of the surface state wave function along the $y$ direction. The remaining $N_{\text{imp}}(1-r/d)$ impurities will not overlap with the surface state wave function and thus will not contribute to $l_{s\Gamma}$. Thus, the matrix element for $N_{\text{imp}}$ impurities scales as

$$|\langle n,\mathbf{k}|V|n',\mathbf{k}'\rangle|^2 \sim \frac{N_{\text{imp}}}{L_x^2 L_z^2 d^2}, \quad (S37)$$

where we note that $r$ is independent of sample dimensions when the latter are large.

Fourth, we recognize that (i) $\sum_{\mathbf{k}}$ and $\sum_{\mathbf{k}'}$ each scale as $L_x L_z$; (ii) $\sum_{n\in\Gamma}$ scales with $d$ (because the number of quantum well states crossing the Fermi energy scales with the film thickness); (iii) $\sum_{n'\in s}$ does not scale with system size (since the number of Fermi arcs does not depend on the film thickness).

Combining the preceding four observations with Eq. (S25), we conclude that

$$\frac{en_s v}{l_{s\Gamma}} \sim \frac{1}{L_x L_z}L_x^2 L_z^2 d \frac{N_{\text{imp}}}{L_x^2 L_z^2 d^2} \sim \frac{N_{\text{imp}}}{L_x L_z d}. \quad (S38)$$

Thus, $1/l_{s\Gamma}$ scales with the volume density of impurities. As mentioned above, these scaling arguments apply for thicker films.

If we repeat the same scaling arguments for $l_{s\bar{s}}$, we find that

$$\frac{en_s v}{l_{s\bar{s}}} \sim \frac{N_{\rm imp}}{L_x L_z d}, \tag{S39}$$

i.e. $1/l_{s\bar{s}}$ is also proportional to the volume impurity density (assuming that those impurities are uniformly distributed in the entire film).

Finally, we can repeat similar arguments for the bulk mean free path. Now the starting point is given by Eqs. (S30) and (S31). Using

$$|\langle n \in {\rm bulk}, \mathbf{k}|V|n' \in {\rm bulk}, \mathbf{k}'\rangle|^2 \sim \frac{N_{\rm imp}}{L_x^2 L_z^2 d^2}, \tag{S40}$$

the outcome reads

$$l_0^{-1} \sim \frac{N_{\rm imp}}{L_x L_z d}, \tag{S41}$$

i.e. the bulk mean free path scales inversely with the volume density of impurities (which is of course well-known).

Thus far, we have assumed that the impurities were randomly but uniformly distributed across the entire film. In the main text, we consider the situation in which the vacancies are concentrated at the surfaces. All of those vacancies contribute to bulk-surface, surface-surface and bulk-bulk scattering. This yields

$$\begin{aligned}
|\langle n \in s, \mathbf{k}|V|n' \in \Gamma, \mathbf{k}'\rangle|^2 &\sim \frac{N_{\rm imp}^{\rm surf}}{L_x^2 L_z^2 d} \to \frac{en_s v}{l_{s\Gamma}} \sim \frac{N_{\rm imp}^{\rm surf}}{L_x L_z} \equiv N \\
|\langle n \in s, \mathbf{k}|V|n' \in \bar{s}, \mathbf{k}'\rangle|^2 &\sim \frac{N_{\rm imp}^{\rm surf}}{L_x^2 L_z^2} \to \frac{en_s v}{l_{s\bar{s}}} \sim \frac{N_{\rm imp}^{\rm surf}}{L_x L_z} \equiv N \\
|\langle n \in {\rm bulk}, \mathbf{k}|V|n' \in {\rm bulk}, \mathbf{k}'\rangle|^2 &\sim \frac{N_{\rm imp}^{\rm surf}}{L_x^2 L_z^2 d^2} \to \frac{1}{l_0} \sim \frac{N_{\rm imp}^{\rm surf}}{L_x L_z d} \equiv \frac{N}{d},
\end{aligned} \tag{S42}$$

where $N$ is the areal density of surface vacancies. Thus, when impurities are localized on the surfaces, the surface-to-surface and bulk-to-surface scattering rates are independent of the film thickness and depend only on the areal density of impurities. However, the bulk mean free path is inversely proportional to the film thickness[S1]. Accordingly, if $d$ is changed while $N$ is kept fixed, then $l_0$ will also change. This change will nevertheless be often unimportant if there is in parallel another source of scattering (as modeled by the factor $\eta$ in the main text), uniformly distributed in the bulk with a volume density that exceeds $N/d$.

---

[S1] It is straightforward to show that this same statement applies to the bulk-to-bulk scattering lengths such as $l_{\Gamma R}$

**Supplementary Figure 1**

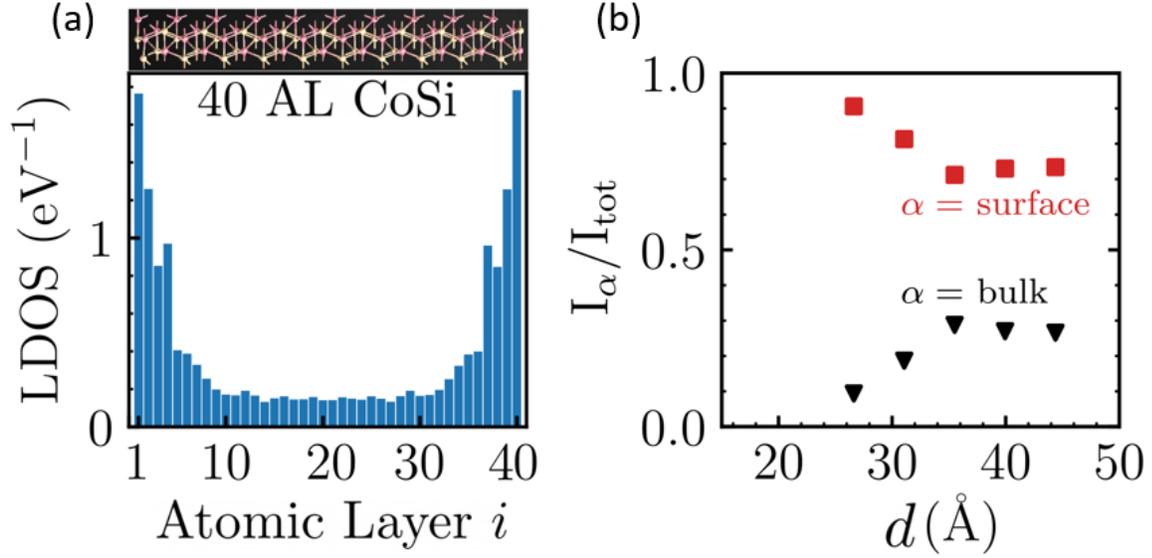

**Participation of Fermi-arc surface states in transport of thin CoSi slabs. (a)** Local density of states (LDOS) at $E_F$ resolved along the thickness of a 40 atomic-layer (AL) pristine CoSi slab with $d = 36.05$ Å, showing strong participation of surface states. LDOS is calculated from the expression

$$(LDOS)_{\text{Atomic Layer } i} = \sum_{j\beta} \frac{1}{(2\pi)^2} \int d^2\mathbf{k}\, A^i_{j\beta}(\mathbf{k}, E_F),$$

where $i$ is the atomic layer number, $j$ is index for the atoms in this atomic layer, $\beta$ is the index for the local atomic orbital on the atoms, and $A^i_{j\beta}(\mathbf{k}, E_F)$ is the corresponding $\mathbf{k}$-resolved spectral weight.
**(b)** Thickness dependence of the fraction of charge current ($I_\alpha/I_{\text{tot}}$) carried by the surface and bulk states respectively, indicating the dominance of surface-state mediated transport.

**Supplementary Figure 2**

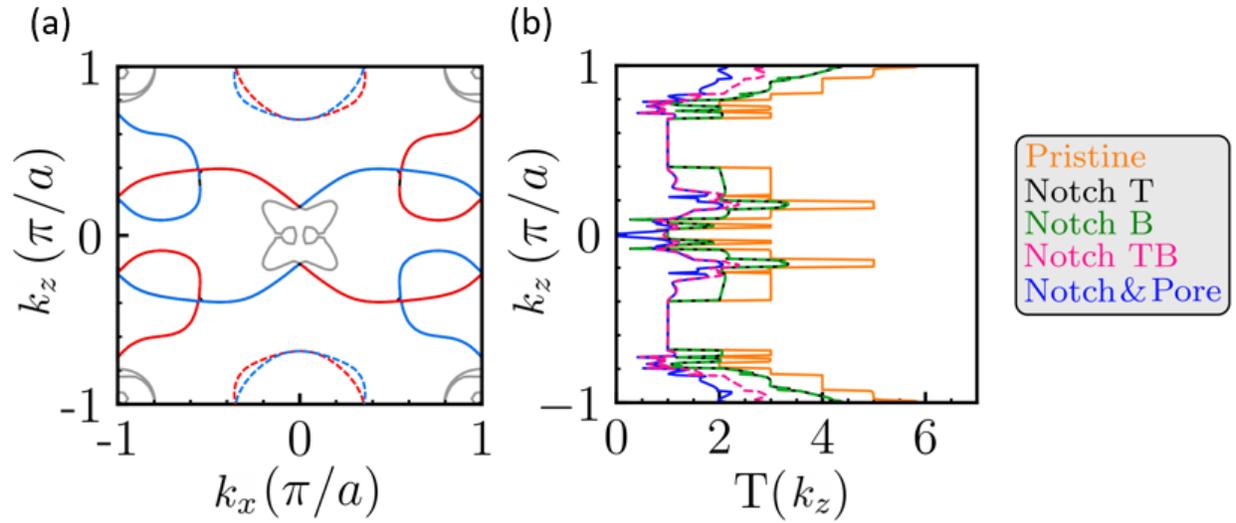

**First-principles based quantum transport of CoSi slabs with 1D line defects. (a)** Fermi surface of a 40AL CoSi slab with $d$ = 36.05 Å. Surface states originated from the top and bottom surfaces are denoted in red and blue lines, respectively. Gray lines indicate the bulk states. **(b)** The $k_z$-resolved transmission T($k_z$) for 40 AL CoSi slabs with various line defects obtained by NEGF calculations using the QuantumATK package. Orange line denotes the ideal film. Black (Notch T) denotes a slab with a notch on the top surface. Green (Notch B) denotes a slab with a notch on the bottom surface. Pink (Notch TB) denotes a slab with a notch on both the top and the bottom surfaces. Blue (Notch & Pore) denotes a slab with a notch on the top surface and a pore through the bulk.

**Supplementary Figure 3**

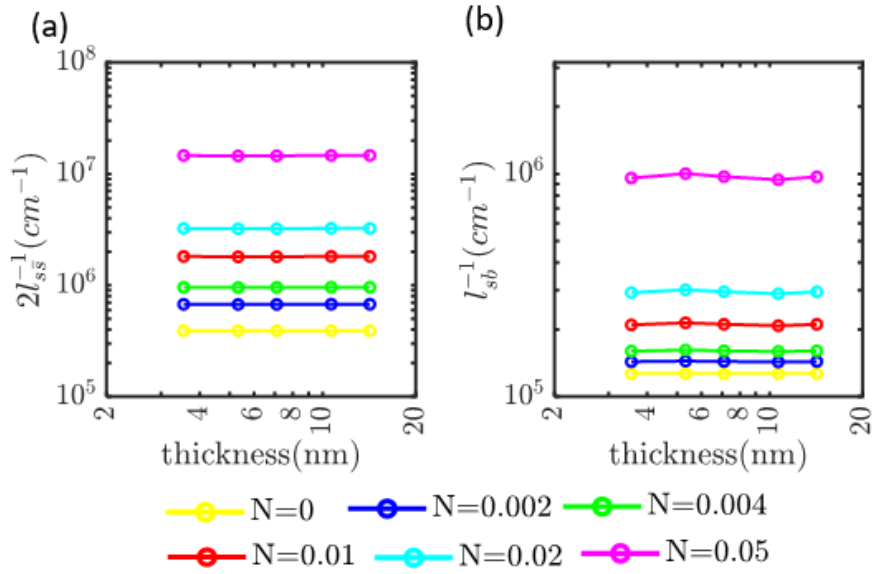

**Scaling of scattering lengths with CoSi slab thickness. (a), (b)** Calculated surface-to-surface ($l_{s\bar{s}}$) and surface-to-bulk ($l_{sb}$) scattering lengths for different values of the surface defect density $N$ (see **Methods**). In the range of thickness corresponding to region $I$ of Fig. 3, all scattering lengths are thickness independent, and the scattering between time-reversed Fermi arcs on the same surface is stronger than surface-to-bulk scattering. The effective surface scattering length $l_s$ in the main text is obtained by: $l_s^{-1} = 2l_{s\bar{s}}^{-1} + l_{sb}^{-1}$.

**Supplementary Figure 4**

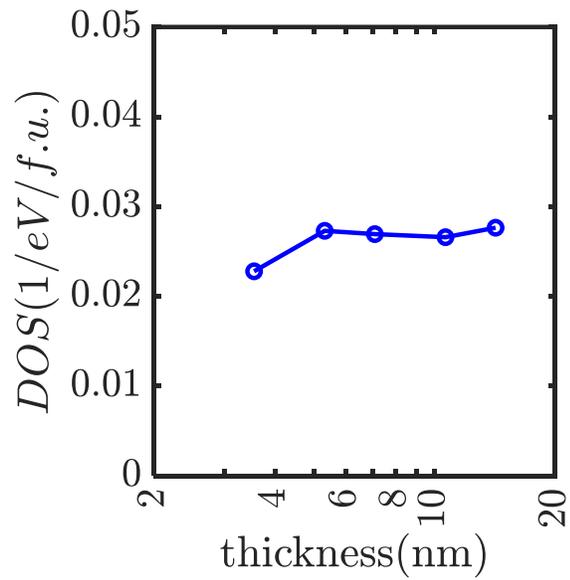

**Scaling of density of bulk states with CoSi slab thickness.** The calculated density of the bulk states per chemical formula unit (f. u.) at the Fermi level hardly varies in the range of thickness corresponding to region *I* of Fig. 3.

**Supplementary Figure 5**

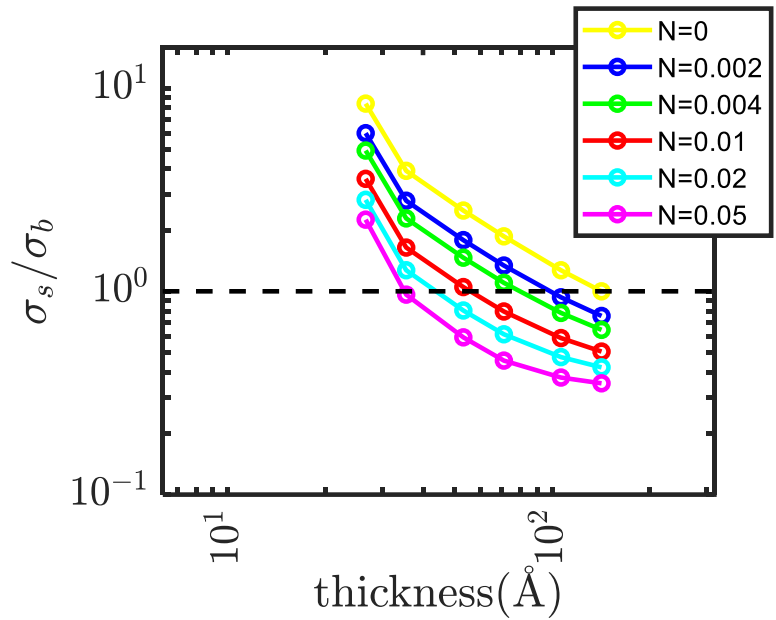

**Scaling of surface-to-bulk conductivity ratio with CoSi thickness**, calculated from first-principles (see **Methods**) for different values of the surface defect density $N$ in the range of thickness corresponding to region *I* of Fig. 3. With increasing $N$, $\sigma_s/\sigma_b$ decreases, indicative of growing dominance of the bulk states over the surface states, giving rise to the sign change in the slope of $\rho/\rho_0$ *vs* $d$ in Figure 3.

**Supplementary Figure 6**

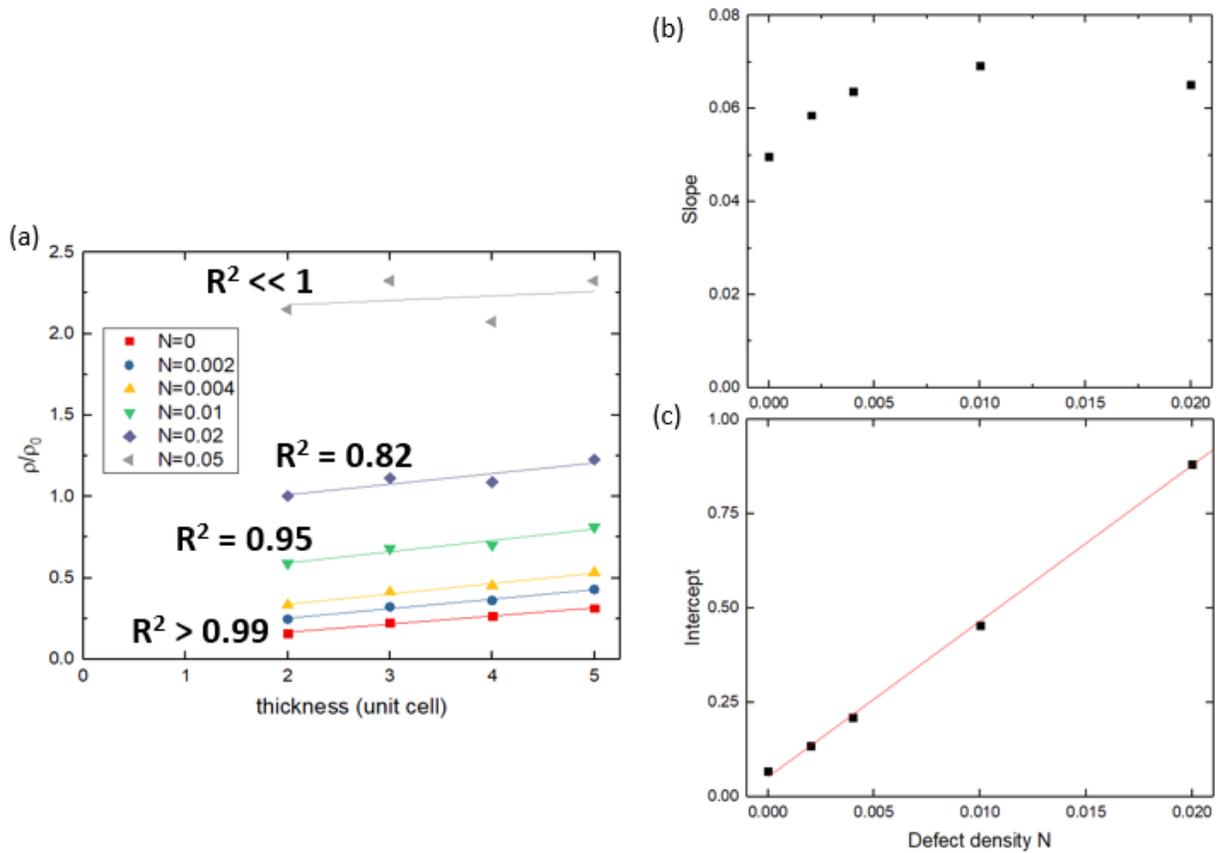

**Scaling of normalized resistivity with CoSi thickness in the ultrathin limit. (a)** Linear fit to the normalized CoSi slab resistivity calculated from first-principles in the thickness range of region *II* (see Fig. 3 in the main text) for different values of the surface defect density $N$. For $N = 0.02$ (equivalent to areal density $\sim 1 \times 10^{13}$ cm$^{-2}$) and below, linear dependence in thickness fits the data well, with a high coefficient of determination $R^2$. **(b)** Slope of the linear fit, which varies weakly with $N$. **(c)** Intercept of the linear fit, which scales linearly with $N$. These findings are consistent with the approximate equation of $\rho/\rho_0 = \sigma_0/\sigma \propto a\,N + b\,d\,\eta$ proposed for region *II* (see main text).

**Supplementary Figure 7**

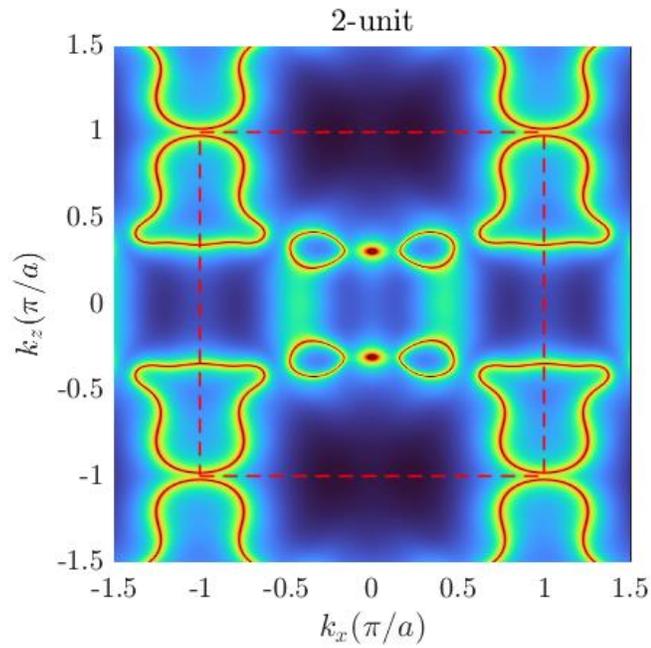

**Spectral weight of 2 unit-cell thick CoSi at the Fermi level.** First-principles calculations reveal the persistent remnants of the Fermi-arc surface states which preserves the conduction down to the ultrathin limit, resulting in decreasing resistivity with decreasing thickness below the critical thickness.